\documentclass[preprint]{aastex}
\usepackage[latin1]{inputenc}
\usepackage{graphicx}
\graphicspath{ {./} }
\usepackage{amsmath}
\usepackage{bm}
\usepackage{mathrsfs}
\usepackage{rotating}
\usepackage{pdflscape}

\slugcomment{}

\shorttitle{}
\shortauthors{Guennou et al.}

\begin{document}

\title{Testing predictors of eruptivity using parametric flux emergence simulations\\}

\author{C. Guennou\altaffilmark{1}, E. Pariat\altaffilmark{1}, J. Leake\altaffilmark{2} and N. Vilmer\altaffilmark{1}}

\affil{\altaffilmark{1} LESIA, Observatoire de Paris, PSL Research University, CNRS, Sorbonne Universit\'es, UPMC Univ. Paris 06, Univ. Paris Diderot, Sorbonne Paris Cit\'e, 5 place Jules Janssen \\92195 Meudon Cedex, France}

\affil{\altaffilmark{2} NASA Goddard Space Flight Center, Greenbelt, MD, USA}

\email{chloe.guennou@obspm.fr}

\begin{abstract}
      
Solar flares and coronal mass ejections (CMEs) are among the most energetic events in the solar system, impacting the near-Earth environment. Flare productivity is empirically known to be correlated with the size and complexity of active regions. Several indicators, based on magnetic-field data from active regions, have been tested for flare forecasting in recent years. None of these indicators, or combinations thereof, have yet demonstrated an unambiguous eruption or flare criterion. Furthermore, numerical simulations have been only barely used to test the predictability of these parameters. In this context, we used the 3D parametric MHD numerical simulations of the self-consistent formation of the flux emergence of a twisted flux tube, inducing the formation of stable and unstable magnetic flux ropes of Leake (2013, 2014). We use these numerical simulations to investigate the eruptive signatures observable in various magnetic scalar parameters and provide highlights on data analysis processing. Time series of 2D photospheric-like magnetograms are used from parametric simulations of stable and unstable flux emergence, to compute a list of about 100 different indicators. This list includes parameters previously used for operational forecasting, physical parameters used for the first time, as well as new quantities specifically developed for this purpose. Our results indicate that only parameters measuring the total non-potentiality of active regions associated with magnetic inversion line properties, such as the Falconer parameters $L_{ss}$, $WL_{ss}$, $L_{sg}$ and $WL_{sg}$, as well as the new current integral $WL_{sc}$ and length $L_{sc}$ parameters, present a significant ability to distinguish the eruptive cases of the model from the non-eruptive cases, possibly indicating that they are promising flare and eruption predictors. A preliminary study about the effect of noise on the detection of the eruptive signatures is also proposed. 
        
\end{abstract}


\clearpage
\section{Introduction}
\label{sec:intro}

Solar flares are sudden brightenings of the solar atmosphere occurring over the whole spectral range, from radio to X-rays, resulting from the release of a huge amount of magnetic energy. Flares can be classified in terms of the strength of the X-ray flux measured close to Earth. They can be divided into two general sub-categories: eruptive flares, accompanied by a coronal mass ejection (CMEs), large ejection of magnetized plasma released into the interplanetary space, and confined flares, for which only electromagnetic radiation is emitted. Both CMEs and solar flares can impact the near-Earth environment and in particular human technologies. The most energetic solar flares can heat and ionize the upper Earth atmosphere, engendering disruption in radio communication and Global Positioning System (GPS) inaccuracies. CMEs can be accompanied by the burst of energetic particles and can be associated with strong geomagnetic storms, inducing ground-level electric fields likely to produce power grids damages. Astronauts, pilots and satellites can also be affected by radiations originating from flares or CMEs. The ability to accurately predict both solar flares and CMEs is therefore fundamental to protect both space and ground-based technologies from strong space weather events. 

The European H2020 research project FLARECAST (Flare Likelihood and Region Eruption Forecasting - \texttt{\url{http://flarecast.eu/}}) aims to develop a fully-automated solar flare forecasting system. FLARECAST will automatically extract various magnetic-field parameters of solar active regions (ARs) from vector solar magnetograms to produce accurate predictions using the state-of-the-art forecasting techniques based on data-mining and machine learning. Various systems of prediction invoking different categories of models have been developed in the past decades, as e.g. the "Theophrastus" tool \citep{McIntosh1990}, the linear-prediction system of \citet{Gallagher2002} used by SolarMonitor, the discriminant analysis of \citet{Barnes2007}, the Automated Solar Activity Prediction (ASAP) of \citet{Colak2009}, based on machine learning and more recently the statistical learning technique of \citet{Yuan2010}. A recent comparison between the current forecasting tools using LOS magnetograms has been performed by~\citet{Barnes2016}, showing that none of them substantially outperformed all others. All these forecasting techniques, including the future FLARECAST forecasting tool, require scalar quantities derived from photospheric magnetic field to be able to make flare and/or eruption predictions. In this paper, we focus on the predictability capabilities of these scalar quantities, in order to determine the most reliable for flare and eruption forecasting use.   

It is now widely accepted that the energy source of solar flares is stored in highly non-potential magnetic fields. The storage of energy, typically estimated in the $10^{28} - 10^{32}$~erg range \citep{Schrijver2012}, arises from a long phase of magnetic stress and free magnetic energy build-up before sudden flare or/and eruption. A large variety of models have been recently developed to explain this storage and released mechanism \citep[see e.g. the recent reviews of][]{Aulanier2014, Janvier2015, Lin2015, Schmieder2015}, but it is still unclear why some ARs trigger a flare, while others remaining quiet. In this context of absence of a clear physical scenario for flare triggering, efforts have been concentrated on empirical flare prediction methods, investigating the behavioral patterns of ARs. 
   
Mapping the 3D coronal magnetic field is not systematically practicable, while the photospheric magnetic field is more easily measurable using spectro-polarimeters such as the Helioseismic and Magnetic Imager \citep[HMI;][]{Schou2012} on board the \textit{Solar Dynamics Observatory (SDO)}. Consequently, flare forecasting tools are mainly based on quantities derived directly from solar magnetograms. As flare occurence is correlated with size, variability and complexity of the ARs, i.e. the associated non-potentiality degree, many attempts have been made in order to find reliable photospheric eruptivity indicators. Global photospheric features, such as magnetic field gradient, currents, magnetic geometry, and magnetic free energy have been proposed to establish a link between photospheric observations and coronal activity. 

However, it has been shown that observing clear pre-signature of flare and eruption in observations is challenging. Many studies investigated the link between changes in some photospheric parameters and the coronal eruptive and flaring activity, by investigating prior temporal changes, performing super-posed epoch analysis or forecasting the likelihood of the flaring events \citep[see e.g][and references therein]{Ahmed2013, Bao1999, Falconer2011, Jing2010, Leka2003a, Mason2010, Schrijver2007, Bobra2016}, with moderate results. More recently, \citet{Algraibah2015} studied the magnetic parameters of about 2000 ARs to search for the best eruptive predictor. They used a wide range of parameters, including a wavelet analysis to resolve multiple-scale changes, and found that that magnetic field properties alone are not sufficient for powerful flare forecasting, although the magnetic-gradient related features appears to be the one of the most reliable indicator for flare predictions. 

Despite all these previous studies, none of the current photospheric indicators, or combinations thereof have yet demonstrated an unambiguous eruptive or flaring criterion. However, controlled cases (e.g., originating from numerical datasets) have barely been used to test the predictability of these parameters. \citet{Kusano2012} presented a parametric analysis of eruption onset, by varying two parameters associated with the magnetic structure. They showed that these two parameters are able to discriminate between eruptive and non-eruptive ARs, although no comprehensive scalar quantity directly measurable is provided. In this work, we use MHD numerical simulations of the formation of stable and unstable magnetic flux ropes \citep{Leake2013, Leake2014} in order to systematically investigate the pre-eruptive signature included in different magnetic parameters. This series of numerical experiments is based on the emergence of a convection zone magnetic flux tube into a solar atmosphere. The interaction of the emerging magnetic flux with the pre-existing magnetic field plays a key role for triggering impulsive events. This class of simulations are thus able to explain and reproduce multiple active solar phenomena~\citep[see e.g. the review of][]{Cheung2014}. Time series of magnetograms from parametric simulations of stable and unstable flux emergence, \textit{i.e. corresponding to respectively quiet AR versus eruptive flare configurations}, are used to compute a large range of parameters. This list includes parameters previously used for operational forecasting, as well as parameters used for the first time for this purpose, such as helicity~\citep{Pariat2017}, the current-weighted magnetic inversion line, and the length of the strong-current magnetic inversion line portions. 
 
The remainder of this paper is as follows. The set of eruptive and non-eruptive simulations are summarized in Section~\ref{sec:simu}. Section~\ref{sec:analysis} describes the magnetograms and the magnetic parameters extracted from our simulation sets. The results are presented in Section~\ref{sec:proxies}, and a discussion on the impact of data masking on the analysis is proposed on Section~\ref{sec:data_masking}. A basic analysis of noise impact on our results is presented in Section \ref{sec:uncer}. Section~\ref{sec:para_MPIL} provides a parametric study of the inversion-line related parameters, and the influence of the different thresholds on eruption predictability. Finally, Section~\ref{sec:summary} summarizes our work and presents our main conclusions.  

\section{Numerical datasets: parametric simulations of eruptive and non-eruptive active regions}
\label{sec:simu}

\subsection{MHD simulations}
\label{subsec:mhd}
In order to test the reliability of photospheric eruptive predictors, we used the three-dimensional visco-resistive magnetohydrodynamic simulations of~\citet{Leake2013, Leake2014}. These simulations describe the partial emergence process of a twisted magnetic flux tube into a stratified solar atmosphere, where a coronal arcade field is present. Both stable and non-stable flux ropes are formed as a result of the flux emergence, depending on the choice of certain parameters. These simulations are thus analogous to respectively quiet ARs and eruptive-flare productive ARs, where the newly formed flux ropes are ejected higher in the solar corona.   

\begin{figure}
\epsscale{0.9}
\plottwo{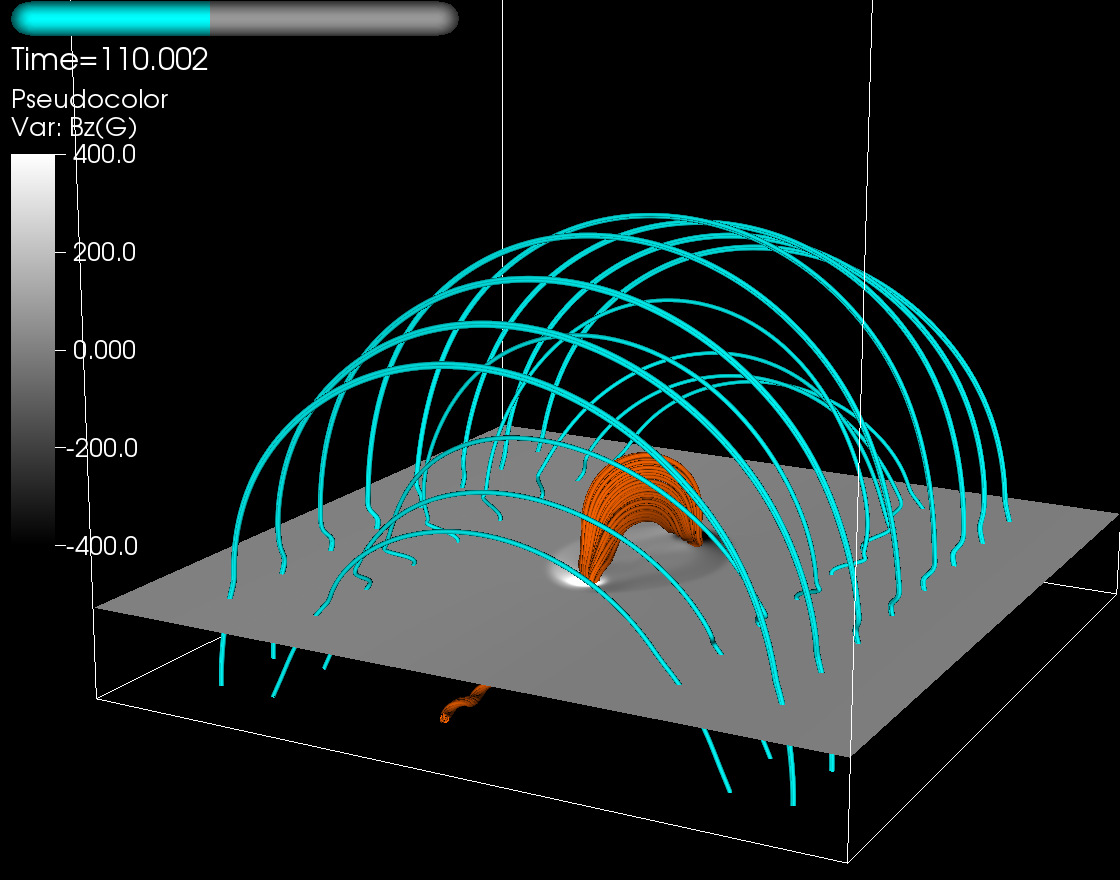}{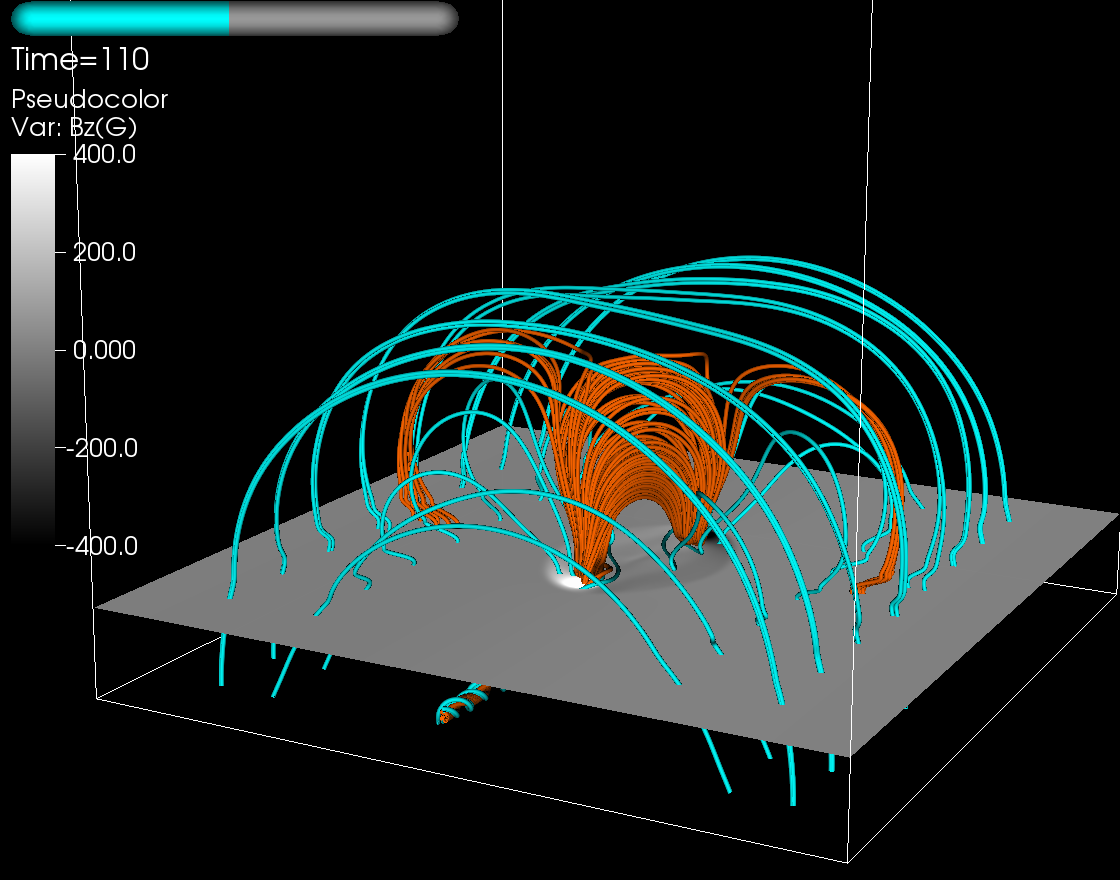}
\caption{\small Emergence of the convection magnetic flux tube into the solar atmosphere at $t = 110\ t_0$ for both the stable SD (left) and the unstable SD E (right) simulations. The cyan lines traces the coronal arcade magnetic field, originating from the lower boundaries whereas the red lines belong to the magnetic flux tube and originates from the $y = \pm \rm{max} y $ side boundaries. The gray scale indicates the magnetic field strength at the surface. \label{fig:fig_1}}
\end{figure}

The evolution of the system is described by the visco-resistive MHD equations (see Equations 1-4 from \citet{Leake2013}), and the plasma is assumed to be fully ionized. The MHD equations are solved using the Lagrangian-remap code Lare3D \citep{Arber2001}, using an irregular cartesian grid. The initial conditions consist of a hydrostatic background atmosphere, stratified such as the solar convection zone, the photosphere/chromosphere, the transition region and the corona. An arcade field covering the entire simulation domain is imposed on this background atmosphere (cyan solid lines on Figure~\ref{fig:fig_1}), and a right-hand twisted flux tube is inserted in the solar convection zone (red solid lines on Figure~\ref{fig:fig_1}), aligned along the $y$ axis. The arcade field is transitionally invariant along the $y$ axis, generated by a source much deeper in the solar convection zone than the initial horizontal flux tube. This background coronal field is designed to reproduce the magnetic field of an old decaying active region. In these simulations, the initially buoyant twisted flux tube, which is line-tied at the side boundaries, partially emerges from the convection zone into the corona due to a perturbation in the flux's tube pressure and density (see Equations 18 and 19 from \citet{Leake2013}) at the center of the flux tube. The later evolution of the emerging field involves surface shearing and rotation, and the formation of a new coronal flux rope above the new active region. The ultimate state of this new flux rope depends on the choice of the overlying field parameters. 

The emergence process of the flux tube is examined through a range of initial coronal arcade strength and orientations, in order to investigate the effects of the external magnetic field on the formation mechanisms of both stable and unstable flux ropes. The flux rope, self-consistently formed by the emergence of the flux tube, is either ejected, i.e. corresponding to an eruptive flare, or confined in the corona, i.e. equivalent to a quiet stable AR, depending on the orientation of the coronal arcade. For the non-eruptive simulations, the direction of the arcade field is the same as that of the top of the flux tube, maximizing the confinement of the flux tube by the arcade field, as shown by the left panel of Figure~\ref{fig:fig_1}. Conversely, the eruptive simulations are driven by the magnetic reconnection between the arcade and the top of the flux tube's axis, due to the opposite orientation of the magnetic field in the two structures. In this case, the horizontal magnetic field $B_x$ changes sign at the interface between the flux tube and the arcade, making this separatrix a favourable location for magnetic reconnection. For each of the two arcade orientations, the arcade magnetic field strength is varied, leading to three different simulations: strong (SD), medium (MD) and weak (WD), corresponding to the initial surface field strength of 26, 19.5 and 13~G. An additional simulation where no coronal field is present (ND) is also performed, resulting in the formation of a stable flux rope.  
  
For the eruptive simulations, the coronal magnetic field strength varies in the same way, namely $\left\lbrace \rm{WD\ E, MD\ E, SD\ E} \right\rbrace $, where E corresponds to "Eruptive" simulations. The continued reconnection process between the emerging field and the arcade starts as the flux rope reaches the corona, changing the connectivity of the system and forming magnetic field lobes on both side of the emerging flux rope. The right panel of Figure~\ref{fig:fig_1} shows the magnetic field lines configuration of the flux tube emergence process for the SD E simulation, at $t = 110\ t_0$ (with $t_0 = 55.7$~s, see Section~\ref{subsec:scaling} for details), slightly before the ejection of the flux rope. The formation of a shearing quadrupole configuration above the photosphere is clearly visible, with a central arcade of the emerging structure (red lines) now allowed to expand higher in the corona, whereas horizontal expansion is limited by the lobes on either sides (in red). Once the central arcade reaches a sufficient height, internal reconnection takes place at about $t \sim 120\ t_0$ (depending on the overlying arcade strength, see \citet{Leake2014} for details) beneath the flux rope axis, and the newly formed flux rope accelerates its vertical expansion, and the flux rope is immediately ejected. The objective of this work is to investigate if some pre-eruptive flare variations in one or more physical parameters are detectable using only photospheric magnetograms, at a reasonable stage before $t \sim 120\ t_0$, providing thus a reliable eruptive indicator.

This set of 7 simulations treats the cases of both eruptive and non eruptive flux rope formation. The presence and the orientation of the coronal arcade are critical parameters for the eruption onset. The ratio of the arcade flux to emerging flux is also a fundamental parameter, controlling the flux rope vertical acceleration, its size and the amount of reconnection allowed to occur.

\subsection{Scaling}
\label{subsec:scaling}

The visco-resistive MHD equations are non-dimensionalized, using a normalisation constant for each variable. In this work, we choose a slightly different normalisation than the original paper, in order to increase the representative size and the magnetic flux of the active regions, to obtain more representative ARs. For the length, we rescale the simulations using the normalizing value $L_0 = 8.5 \times 10^5$~m, corresponding to 5 times the original value adopted by \citet[][see Section 2.2]{Leake2013, Leake2014}. The two others normalizing values, the magnetic field ($B_0 = 1300$~G) and gravitational acceleration ($g_0 = g_{\rm{sun}} = 274$~m s$^{-2}$) remain unchanged. The other normalizing variables that are affected by this rescaling are namely the density ($\rho_0 = 5.77 \times 10^{-5}$~kg m$^{-3}$),  the velocity ($v_0 = 15.26 \times 10^3$~m s$^{-1}$), the time ($t_0 =55.7 $~s), the temperature ($T_0 = 28.2 \times 10^3$~K), the current density ($j_0 = 0.122$~A/m$^{2}$), the viscosity ($\nu_0 =7.49 \times 10^5 $~kg m$^{-1}$ s$^{-1}$) and the resistivity ($\eta_0 = 16.3 \times 10^3\ \rm{\Omega}$m). The Reynold's number and the magnetic Reynolds number are kept in the same order as in the initial simulations, both evaluated to 100.

Using a larger spatial scaling than that of the initial analysis allows us to obtain larger active regions and larger magnetic flux, therefore more representative of the observed eruptive active region. Still, we are limited to the study of a relatively small active region with a characteristic flux of $1-2 \times 10^{21}$~Mx during the early emergence process, and a characteristic size of about 30~Mm, roughly giving an area of about $296\ \mu$hs (i.e. microhemisphere - 1$\mu$hs $= 3.04 \times 10^6$~km$^2$). According to \citet{Sammis2000}, our simulations belongs to the smallest flaring active region. For comparison, the authors have shown that during the 1989-1997 time frame, all the X4 flares only originate from active regions with an area greater than $1000\ \mu$hs. It should be noted that not all active regions in this size range are flaring: size seems to be a necessary condition to X4 flares, but not sufficient. Given our framework, our study is thus limited in terms of size and complexity of active regions that we can not explore: the reliability of eruptive indicators is tested only for a given class of active region size and complexity. Nonetheless, it is worth noting that a recent study by~\citet{Toriumi2017} concludes that a complex $\delta$-sunspots is not a necessary condition for flaring ARs, even for the X-class flares. Thus, this kind of controlled study is important in the sense that the size and the complexity of an active region are not a discriminant parameter for eruptivity: small active regions can still produce flares, while two active regions in the same size and complexity range do not necessarily have the same flaring likelihood.
    
However, if the rescaling of the simulations set enable us to have greater size, it also influences other quantities used in the simulations. Using our scaling, the transition region is 8.5~Mm thick (1.7~Mm in the original simulations), which is significantly higher than its characteristic size of about 0.1~Mm. However, given the spatial resolution imposed by the total domain size and the computational cost, such a thickness is required to resolve the large temperature and density gradient occurring in this part of the solar atmosphere.   

\section{Analysis: extracting the physical properties of ARs}
\label{sec:analysis}

In the following, the methodology that we follow to investigate the reliability of the eruptive flare indicators is detailed. From the series of 2D-plane vector magnetograms of the 3 eruptive and 4 stable simulations, we compute a series of parameters in order to detect some particular behavior associated only with the eruptive simulations. A predictor can be considered as reliable if it shows significant change(s) in a reasonable timescale prior to the eruption. Since flares and eruptions are driven by a store-and-release mechanism of magnetic energy, eruptive indicators must provide indications of such process occurring. Such signature includes threshold beyond which the system becomes unstable, most likely generating an eruption or a flare. Therefore a reliable predictor should present a different behavior respective to the eruptive or the non-eruptive nature of the simulations, but a similar behavior for simulations of the same nature. This signature should also be significant enough to be measured, and detectable in a sufficient time prior to the eruption to be able to then perform operational forecasting. Here, we focus only on determining which quantities could potentially be proficient eruptive flare predictors, i.e. parameters associated with higher values for the 3 eruptive simulations $\left\lbrace \rm{WD\ E, MD\ E, SD\ E} \right\rbrace $ than that of the stable $\left\lbrace \rm{ND, WD, MD, SD} \right\rbrace $ prior to the eruption. The associated thresholds needed for performing operational predictions require the use of (1) real photospheric magnetogram observations and (2) substantially more ARs sample to be statistically significant.    

\subsection{Magnetograms}
\label{subsec:magneto}

From the time series of the 3D MHD simulations, we first extract photospheric-like 2D-plane vector magnetograms, by interpolating the cube domain at $z=0$ \citep[as defined by][]{Leake2013}, onto a regular grid using a resolution of 0.86~$L_0$ (730~km/pixel) in both directions, which is equivalent to the instrument resolution of about 1''. This resolution is the same as that of the \textit{SDO}/HMI magnetograms. In our simulations, the surface ($z = 0$) is defined as the beginning of the temperature minimum region, i.e. corresponding to the region of the lowest pressure scale height. Since the eruptive flare occurs around $t \sim 120 t_0$ for the three eruptive simulations $\left\lbrace \rm{WD\ E, MD\ E, SD\ E} \right\rbrace $, we restrict the time windows of our analysis from $t = 0\ t_0$ to $t = 150\ t_0$, using a time sampling of $\Delta t = 5\ t_0$, where $t_0 = 55.7 s$. Therefore, we finally obtain 31 magnetograms series for each vector magnetic field component $B_x, B_y, B_z$, and for each of the 7 parametric numerical simulations $\left\lbrace \rm{ND, WD, WD\ E, MD, MD\ E, SD, SD\ E} \right\rbrace $.

We first apply a  mask to the entire magnetograms series, excluding the pixels for which $B  = \sqrt{B_x^2 + B_y^2 + B_z^2} < B_{mask}$, where $B_{mask} = 30$~G. This $B_{mask}$ threshold has been chosen in order to be greater than the background magnetic field due to the coronal arcade field, respectively about 26, 19.5 and 13~G for the SD, MD and WD simulations (see Section~\ref{subsec:mhd}). In observational analysis, a threshold is usually applied to exclude noisy data and only retain strong field areas. Our data mask threshold is relatively low compared to the typical uncertainties measured in observed magnetograms and so far the usual threshold applied to data. This is due to the relatively small typical values of magnetic field in our simulations, with $B_z$ values in the [-800~G,800~G] range, due to the small size and flux of our ARs (See Section~\ref{subsec:scaling}). Many different masking methods have been used in previous analyses, using different threshold values applied on either $B$ or $B_z$ magnetograms, and will be briefly introduced in Section \ref{sec:data_masking}. The impact of applying a data mask on the detection of photospheric eruptive signatures will also be discussed in more detail in Section \ref{sec:data_masking}. The effects of noise, using a random perturbation of the magnetograms, will also be investigated in Section~{\ref{sec:uncer}}.

\begin{figure}
   \centering
   \includegraphics[width=0.65\columnwidth]{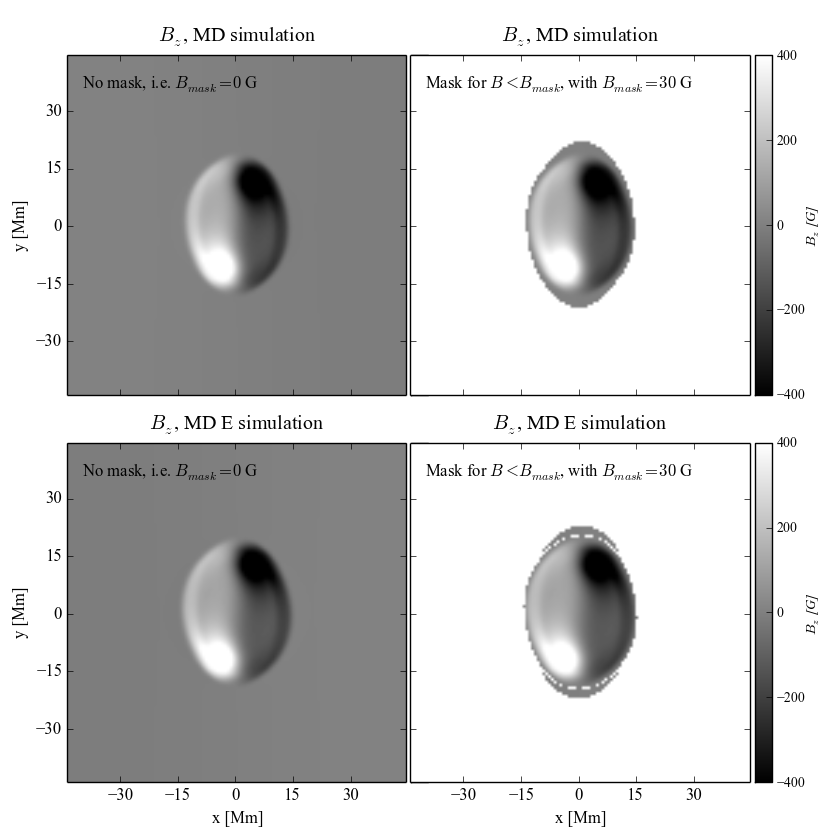}
   \caption{\small Magnetograms of the $B_z$ component at $t=100\ t_0$ for both the MD (top) and MD E (bottom) simulations. Both non-masking (left) and masking (right) magnetograms are displayed, in order to show the reduction of the region of interest for further analysis. Eruptive and non-eruptive simulations present very similar magnetograms, except near the external edge of the polarities, due to the quadrupolar geometry above the surface. \label{fig:fig_2}}

\end{figure}

Figure \ref{fig:fig_2} shows an example of such magnetograms, for both eruptive and non-eruptive simulations. Top panels display $B_z$ magnetograms for the MD simulation, either masking (right) or not (left) of the data, while bottom panels exhibit the same photospheric maps for the MD E simulations. Masking data has a great impact on the area of the region of interest, used then for the computation of the different eruptive indicators. Eruptive and non-eruptive magnetograms are very similar, except near the external edge of the polarities. This is due to the quadrupolar nature of the eruptive simulations: as mentioned before, the coronal arcade has an opposite orientation with respect to the flux tube, i.e the arcade $B_x$ component has the opposite sign, to favour reconnection above the flux tube axis. From such time series of masked magnetograms, we compute a set of scalar parameters, most of them typically used in standard solar flare operational forecasting methods.

\subsection{The parameters}
\label{sec:parameters}

All the quantities considered in this work are scalar and can be derived using exclusively vector magnetic field data. They described the overall physical condition of an active region and the ongoing evolution of its magnetic field. Most of them have been chosen based on their previous identification as potential flare predictors, but new quantities such as helicity or $WL_{sc}$, the current integral along the magnetic polarity inversion line, are also tested. Each derived quantity is parametrized such as one single number is able to characterize the state of the whole active region at a given time. For quantities that are spatially distributed, such as $B_z(x,y)$, we compute the four first moments as described in \citet{Leka2003a}: the mean, the standard deviation, the skewness and the excess kurtosis (i.e. the kurtosis $-3$, for comparison to normal distribution). The skewness describes the asymmetry of the spatial distribution, while the kurtosis accounts for small changes in extremal values. As pointed out by \citet{Leka2003a}, computing the four first moments of the spatial distribution allows us to quantify the subtle evolution of the different parameters.       

\subsubsection{Magnetic field}
\label{subsubsec:B_prop}

This first class of parameters is directly related to magnetic field evolution. All these parameters, listed in Table~\ref{Tab:magn_pro}, come from the work of \citet[][, hereafter LB03]{Leka2003a}, and readers are referred to Section 3 of their paper for a full detailed description. The spatial/scalar column specifies if the computed parameters are spatially distributed or not. If so, the four first moments are computed, easily identifiable by the $\mathscr{M}$ notation. The temporal evolution of the total, vertical and horizontal magnetic fields, respectively noted $B$, $B_z$, $B_h$, and their associated moments provide insight about how the emerging field changes the distribution of $\bm{B}$, and the overall polarity imbalance of the active region. The associated horizontal spatial gradients $\nabla_h B$, $\nabla_h B_h$, $\nabla_h B_z$, and the corresponding moments indicate how much the magnetic field is sheared and distorted, an important indicator of the AR non-potentiality, especially along the neutral line. We also compute 4 quantities related to the AR flux: the total unsigned flux $\phi_{tot}$, the net flux $\phi_{net}$, and the positive and negative flux $\Phi_{\pm}$, providing insights on the overall magnetic flux and the flux imbalance during the partial emergence process. Finally, we compute the photospheric free magnetic energy, which refers to the energy directly available for eruptive activity, i.e. measured with respect to the potential field energy (see the corresponding equation in Table~\ref{Tab:magn_pro}). It is worth noting that in our case, the free magnetic energy is computed from surface integrals, not from the whole AR volume, which would require 3D magnetic field extrapolations.

\begin{table}
\caption{Parameters relying on magnetic field tested in the present study.}             
\label{Tab:magn_pro}      
\centering                          
\begin{tabular}{c c c c}        
\hline\hline                 
Description & Formula & Predictor type & Reference \\
\hline
Total magnetic field & $B = \sqrt{B_x^2 + B_y^2 + B_z^2}$ & $\mathscr{M}[B]$ & \citet{Leka2003a} \\[7pt]
Normal magnetic field & $B_z$ &  $\mathscr{M}[B_z]$ & \citet{Leka2003a} \\[7pt]
Total horizontal magnetic field & $B_h = \sqrt{B_x^2 + B_y^2}$ & $\mathscr{M}[B_h]$ & \citet{Leka2003a} \\[7pt]
Horizontal gradient of $B$ & $\nabla_h B = \sqrt{\frac{\partial B}{\partial x}^2  + \frac{\partial B}{\partial y}^2}$ & $\mathscr{M}[\nabla_h B]$ & \citet{Leka2003a} \\[7pt]
Horizontal gradient of $B_z$ & $\nabla_h B_z = \sqrt{\frac{\partial B_z}{\partial x}^2  + \frac{\partial B_z}{\partial y}^2}$ & $\mathscr{M}[\nabla_h B_z]$ & \citet{Leka2003a} \\[7pt]
Horizontal gradient of $B_h$ & $\nabla_h B_h = \sqrt{\frac{\partial B_h}{\partial x}^2  + \frac{\partial B_h}{\partial y}^2}$ & $\mathscr{M}[\nabla_h B_h]$ & \citet{Leka2003a} \\[7pt]
Total unsigned flux & $\Phi_{tot} = \sum |B_z| \delta x \delta y$ & $\Phi_{tot}$ & \citet{Leka2003a} \\[7pt]
Total net flux & $\Phi_{net} = |\sum B_z \delta x \delta y |$ & $\Phi_{net}$ & \citet{Leka2003a} \\[7pt]
Positive/negative flux & $\Phi_{\pm} = \sum\limits_{\pm B_z > 0} B_z \delta x \delta y $ & $\Phi_{\pm}$ & \citet{Leka2003a} \\[10pt]
Free magnetic energy & $\rho_e = \frac{(\bm{B^{pot} - B})^2}{2 \mu_0}$ & $\mathscr{M}[\rho_e]$ & \citet{Leka2003a} \\[10pt]
Total free magnetic energy & $E_e = \sum \rho_e \delta x \delta y$ & $E_e$ & \citet{Leka2003a} \\[7pt] 
\hline                      
\end{tabular}
\end{table}

\subsubsection{Magnetic field geometry}

This second class of parameters, listed in Table~\ref{Tab:magn_geom}, quantifies the morphology of the AR magnetic field and its deviation from the potential configuration. The inclination angle $\gamma$ measures how much a magnetic field is inclined relative to the vertical axis. It reflects the magnetic field orientation, providing hints about the flux emergence evolution. In the present work, we adopt the following definition
\begin{equation}
\gamma(x,y) = \tan^{-1}\left(\frac{B_h(x,y)}{|B_z(x,y)|}\right).
\end{equation}
As defined, the inclination angle tends towards small values when the magnetic field is approaching the vertical, while great values correspond to an almost horizontal magnetic field. This quantity indicates the morphology evolution of the magnetic field during the emerging process.

In a force-free field framework, the Lorentz force is null and the magnetic field follows
\begin{equation}
\nabla \times \bm{B} = \alpha \bm{B},
\end{equation}
where $\alpha$ is the local twist parameter, referring to the torsion of each individual field line. Using magnetogram data, $\alpha$ can be approximated by \citep[see e.g.][]{Pevtsov1994}
\begin{equation}
\alpha(x,y) = \frac{\nabla_h \times B_h(x,y)}{B_z(x,y)} = \mu_0 \frac{J_z(x,y)}{B_z(x,y)},
\end{equation}
where $\mu_0$ is the permeability in free space. Using a set of 133 of flaring ARs, \citet{Nindos2004} found that the preflare value of the twist parameter $\alpha$ was in general higher for M-class eruptive flares, suggesting the twist parameter as a reliable indicator for CMEs. 

The magnetic shear angle has been extensively used in previous studies to measure the non-potentiality of active regions. Initially introduced by \citet{Lu1993} and \citet{Hagyard1984}, the three dimensional shear angle $\Psi$ measures the angle between the observed magnetic field and its potential component, while its horizontal projection, namely the "horizontal" or "planar" shear angle $\psi$ quantifies the azimuthal difference between the observed and potential magnetic field. The potential magnetic field is computed following the method of \citep{Valori2012} We also tested the additional shear angle-related parameters $A[\Psi > 80^{\circ}]$ and $A[\psi > 80^{\circ}]$ proposed by \citet{Leka2003a}, corresponding to the AR area where respectively the 3D and the horizontal shear angle exceeds $80^{\circ}$. 

\begin{table}
\caption{Parameters relying on magnetic field geometry tested in the present study.}             
\label{Tab:magn_geom}      
\centering                          
\begin{tabular}{c c c c}        
\hline\hline                 
Description & Formula & Predictor type & Reference \\
\hline
Inclination angle & $\gamma = \tan^{-1}(\frac{B_h}{|B_z|})$ & $\mathscr{M}[\gamma]$ & \citet{Leka2003a} \\[9pt]
Twist parameter & $\alpha = \mu_0 \frac{J_z}{B_z}$ & $\mathscr{M}[\alpha]$ & \citet{Leka2003a} \\[7pt]
Horizontal shear angle & $\psi = \cos^{-1}\left[\frac{\bm{B_h^{pot} \cdot B_h}}{\bm{B_h^{pot} B_h}}\right]$ & $\mathscr{M}[\psi]$ & \citet{Leka2003a} \\[12pt]
3D shear angle & $\Psi = \cos^{-1}\left[\frac{\bm{B^{pot}\cdot B}}{\bm{B^{pot}B}}\right]$ & $\mathscr{M}[\Psi]$ & \citet{Leka2003a} \\[12pt]
High shear angle area & $A[\Psi > 80^{\circ}] = \sum\limits_{\Psi > 80^{\circ}} \delta x \delta y$ & $A[\Psi > 80^{\circ}]$ & \citet{Leka2003a} \\[9pt]
High 3D shear angle area & $A[\psi > 80^{\circ}] = \sum\limits_{\psi > 80^{\circ}} \delta x \delta y$ & $A[\psi > 80^{\circ}]$ & \citet{Leka2003a} \\[9pt]
Magnetic helicity flux & $\dot{H}_m = \int G_{\theta}(\bm{x}) B_h \rm{d}S$ & $\dot{H}_m$ & \citet{Pariat2005} \\[9pt]
\hline                      
\end{tabular}
\end{table}

Magnetic helicity measures to what extent the magnetic field lines are wrapped around each other, and how much the individual magnetic field lines are twisted and writhed, relatively to their lowest energy state. This parameter provides a quantitative estimation of the geometric properties of the magnetic field lines. Because helicity is a conserved MHD quantity, even in resistive MHD where the dissipation is very small \citep{Pariat2015}, important efforts have been carried out concerning its estimation and its relation with solar flares \citep[see e.g.][]{Nindos2004, Demoulin2007, Demoulin2009, Park2010}. In the present study, helicity has been computed using the $G_{\theta}$ proxy for the helicity flux density from \citet{Pariat2005}, using the series of photospheric magnetograms. In this framework, the time variation of the relative magnetic helicity can be written as

\begin{equation}
\dot{H}_m = \int_{\mathcal{S}} - \frac{B_z}{2 \pi} \int_{\mathcal{S'}} \frac{1}{r^2} (\bm{r} \times (\bm{u} - \bm{u'}))_n\ B_z'\ d\mathcal{S'} d\mathcal{S},
\end{equation}
where $\mathcal{S}$ is the photospheric surface, $B_z$ is the magnetic field normal to this surface, $\bm{r} = \bm{x} - \bm{x'}$ is the vector between the two photospheric position $\bm{x}$ and $\bm{x'}$, with $\bm{u}$ and $\bm{u'}$ the associated flux transport velocity of the photospheric field line footpoints. While the flux transport velocity could be directly extracted from the simulations, the velocity $\bm{u}$ has been estimated using the differential affine velocity estimator for vector magnetograms (DAVE4VM) from \citet{Schuck2008} in order to analyze the simulation data as observations. Using simulations, \citet{Welsch2007} have tested the robustness of the DAVE algorithm, the line-of-sight magnetogram equivalent of the DAVE4VM algorithm \citep{Schuck2008}, and found a good agreement between the simulated and the DAVE-derived velocities. 

\subsubsection{Current properties}

Current-carrying magnetic fields are understood to be the building block for understanding the flares and CMEs drivers. The current properties computed here (see Table~\ref{Tab:current_para}) allow us to quantify the energy stored by the magnetic field relative to its lowest energy state given its boundary distributions, i.e. potential. The vertical current density component $J_z$ and the associated moments can be estimated through the classical Amp\`{e}re's law. The total, net, positive and negative currents, respectively noted as $I_{tot}$, $I_{net}$ and $I_{\pm}$, are estimated using various reckoning through the AR area. The heterogeneity and chirality current density components are also derived, according to \citet{Zhang2001}

\begin{equation}
\label{eq:zhang}
\bm{J}(x,y) = \frac{B}{\mu_0}\nabla \times \bm{b} + \frac{1}{\mu_0}\nabla B \times \bm{b},
\end{equation}  
with $\bm{B} = B \bm{b}$. The first term of Equation~\ref{eq:zhang} refers to the current of chirality and the second relates to the heterogeneity, perpendicular by construction to the magnetic field. Thus, in the case of the heterogeneity current dominates over the chirality component, the AR magnetic field is far from the force-free field hypothesis. The associated four moments for each components are also evaluated, as well as the total and net heterogeneity and chirality currents, $I_{tot}^{h, ch}$, $I_{net}^{h, ch}$. We also computed the net current originating from each polarity, $I_{net}^B$, as described in \citet{Leka2003a}
\begin{equation}
|I_{net}^B| = |\sum J_z(B_z^{+}) \delta x \delta y| + |\sum J_z(B_z^{-}) \delta x \delta y|,
\end{equation}
as well as the vertical contribution to the current helicity density $h_c$ along the vertical axis. Indeed, the current helicity density is defined as $\bm{B} \cdot \bm{J}$, but since only the vertical current component $J_z$ can be deduced from the observations, the current helicity computed here is thus limited to the vertical contribution alone. From this partial estimation of the helicity density, we then derived the total and net partial helicity, $H_c^{tot}$ and $H_c^{net}$, characterizing the current helicity imbalance over the ARs. \citet{Bao1999} examined the relation between flare activity and the vertical contribution of the current helicity, and found that the time variation of partial current helicity is higher in flaring in ARs, suggesting partial and most probably total current helicity as a valuable eruptive indicator. 

Determining whether or not the net electric current $I_{net}^{\pm}$ is neutralized over individual polarities of ARs is crucial for some theoretical flare and CMEs models~\citep[e.g.][]{Melrose1991, Parker1996, Titov1999, Forbes2010}. If the AR currents are fully neutralized, the net current integrated over one photospheric polarity is set to zero. Recent observations and simulations tend to confirm the existence of non-neutralized ARs \citep[e.g.][and references therein]{Torok2014, Dalmasse2015}. \citet{Forbes2010} argued that neutralized ARs may inhibit the eruption process. Hence, the measurement of direct and return current within ARs could provide insight about the flaring activity. The computations of both the direct current $I_d$ and the return current $I_r$ have been carried out following \citet{Torok2014}, who used the same simulations as this present work. They carried on the computation at the base of the corona, while we only concentrated on surface, i.e. photospheric measurements. Several combinations of $I_d$ and $I_r$ have also been tested: $I_d + I_r$, $|I_d/I_r|$ as well as the same quantities normalized by the total flux $\Phi_{tot}$ of the AR (see Section \ref{subsubsec:B_prop}). 

\begin{landscape}
\begin{table}
\caption{Current properties tested in the present study.}             
\label{Tab:current_para}      
\centering                          
\begin{tabular}{c c c c}        
\hline\hline                 
Description & Formula & Predictor type & Reference \\
\hline
Current density & $J_z = \frac{1}{\mu_0} \left(\frac{\partial B_y}{\partial x} - \frac{\partial B_x}{\partial y}\right)$ & $\mathscr{M}[J_z(s)]$ &  \citet{Leka2003a} \\[10pt]
Chirality current density & $J_z^{ch} = \frac{B}{\mu_0} \left(\frac{\partial b_y}{\partial x}  - \frac{\partial b_x}{\partial y}\right)$, with $\bm{B} = B\bm{b}$ & $\mathscr{M}[J_z^{ch}]$ & \citet{Zhang2001} \\[10pt]
Heterogeneity current density & $J_z^{h} = \frac{1}{\mu_0} \left(b_y\frac{\partial B}{\partial x}  - b_x \frac{\partial B}{\partial y}\right)$, with $\bm{B} = B\bm{b}$ & $\mathscr{M}[J_z^{h}]$ & \citet{Zhang2001} \\[10pt]
Total current & $I_{tot} = \sum |J_z| \delta x \delta y$ & $I_{tot}$ & \citet{Leka2003a} \\[12pt]
Net current & $I_{net} = \sum J_z \delta x \delta y$ & $I_{net}$ & \citet{Leka2003a} \\[7pt]
Positive/negative current & $I_{\pm} = \sum\limits_{J_z >(<) 0} J_z \delta x \delta y$ & $I_{\pm}$ & \citet{Leka2003a} \\[7pt]
Total heterogeneity/chirality currents & $I_{tot} = \sum |J_z| \delta x \delta y$ & $I_{tot}^{h, ch}$ & \citet{Leka2003a} \\[12pt]
Net heterogeneity/chirality currents & $I_{net} = \sum J_z \delta x \delta y$ & $I_{net}^{h, ch}$ & \citet{Leka2003a} \\[7pt]
Net current from each polarity & $|I_{net}^B| = |\sum\limits_{B_z > 0} J_z \delta x \delta y| + |\sum\limits_{B_z < 0} J_z \delta x \delta y| $ & $|I_{net}^{B}|$ & \citet{Leka2003a} \\[9pt]
Vertical contribution to current helicity density & $h_c = B_z \left( \frac{\partial B_y}{\partial x} - \frac{\partial B_x}{\partial y}\right)$ & $\mathscr{M}[h_c]$ & \citet{Leka2003a} \\[9pt]
Total partial current helicity & $H_c^{tot} = \sum |h_c| \delta x \delta y$ & $H_c^{tot}$ & \citet{Leka2003a} \\[9pt]
Net partial current helicity & $H_c^{net} = |\sum h_c \delta x \delta y|$ & $H_c^{net}$ & \citet{Leka2003a} \\[9pt]
Direct current in the positive polarity & $I_{d} = \sum\limits_{B_z > 0}Jz^{+} \delta x \delta y$ & $I_{d}$ & \citet{Torok2014} \\[9pt]
Return current in the positive polarity & $I_{r} = \sum\limits_{B_z > 0}Jz^{-} \delta x \delta y$ & $I_{r}$ & \citet{Torok2014} \\[9pt]
\hline                      
\end{tabular}
\end{table}
\end{landscape}

\subsubsection{Lorentz forces}

The Lorentz forces are thought to be an important parameter for solar eruptions. In particular, \citet{Fisher2012} argued that in eruptive-flare ARs, the impulse coming from the Lorentz forces is dominant over all other forces, suggesting that an observed change in the Lorentz force before the eruption could be a pre-eruptive-flare signature. To test this proposition in our simulations, we used the expressions derived by \citet{Fisher2012} in order to compute the horizontal and vertical components $F_x$, $F_y$, $F_z$ of the Lorentz forces. We also estimated the total Lorentz force $F$ and the normalized horizontal and vertical components $\tilde{F_x}$, $\tilde{F_y}$, $\tilde{F_z}$.

\begin{table}
\caption{Lorentz forces parameters tested in the present study.}             
\label{Tab:lorentz_para}      
\centering                          
\begin{tabular}{c c c c}        
\hline\hline                 
Description & Formula & Predictor type & Reference \\
\hline
Total Lorentz force & $F = \sum B^2 \delta x \delta y$ & $F$ & \citet{Fisher2012} \\[7pt]
$x$-component of the Lorentz force & $F_x = \frac{1}{\mu_0}\sum B_x B_z \delta x \delta y$ & $F_x$ & \citet{Fisher2012} \\[7pt]
$y$-component of the Lorentz force & $F_y = \frac{1}{\mu_0}\sum B_y B_z \delta x \delta y$ & $F_y$ & \citet{Fisher2012} \\[7pt]
$z$-component of the Lorentz force & $F_z =  \frac{1}{2 \mu_0}\sum (B_x^2 + B_y^2 + B_z^2) \delta x \delta y$ & $F_z$ & \citet{Fisher2012} \\[7pt]
Normalized $F_x$ & $\tilde{F_x} = F_x/F$ & $\tilde{F_x}$ & \citet{Fisher2012} \\[7pt]
Normalized $F_y$ & $\tilde{F_y} = F_y/F$ & $\tilde{F_y}$ & \citet{Fisher2012} \\[7pt]
Normalized $F_z$ & $\tilde{F_z} = F_z/F$ & $\tilde{F_x}$ & \citet{Fisher2012} \\[7pt]
\hline                      
\end{tabular}
\end{table}

\subsubsection{Magnetic Polarity Inversion Line (MPIL) properties}
\label{subsubsec:MPIL_prop}

Within an AR, the magnetic field is sheared, stressed and twisted, thus deviating from the potential state of minimum energy. In particular, near the magnetic polarity inversion line (MPIL), the magnetic field can be almost perpendicular to the potential field \citep[see e.g.][]{Hagyard1990, Falconer2003}. This region of strong magnetic shear is particularly well-observed in X-rays, corresponding to the sigmoidal structures with an overall shape of ``S`` or an inverse ``S`` for bipolar ARs. Evaluating the properties of the magnetic field and currents near the MPIL provides a quantitative measure of the AR's overall non-potentiality, even though some studies such as \citet{Li2005}, did not found obvious strongly sheared magnetic neutral line or strong overlying arcade associated with a X3 flare event. Since the degree of non-potentiality seems correlated with eruptive flare productivity \citep{Falconer2002}, the MPIL and near-MPIL areas properties are potentially proficient predictors. 

We computed a series of parameters relying on MPIL lengths, most of them introduced by~\citet{Falconer2003, Falconer2006, Falconer2008}. The total length of the strong field neutral line $L_s$ is defined as the length of the MPIL where $B_h^{pot}$ is greater than a given threshold $B_{h}^{th}$. The quantity $L_{ss}$ defined the length of the MPIL for which the observed horizontal magnetic field $B_h$ is greater than the given threshold $B_{h}^{th}$ and the shear angle is greater than $\psi_{th}$. The total strong-gradient length of the MPIL $L_{sg}$ is specified for MPIL regions where $B_h^{pot}$ is greater than the previous defined threshold $B_{h}^{th}$ and where the horizontal gradient of the vertical magnetic field $\nabla_h B_z > \nabla_h B_z^{th}$. We introduce here a new measure of the AR non-potentiality through the total strong-current length $L_{sc}$, measured along the MPILs, where $B_h > B_{h}^{th}$ and the current density $J_z$ is greater than a given threshold $J_z^{th}$. 

The quantities $WL_{ss}, WL_{sg}$ were first introduced by \citet{Falconer2008} while we define the new parameter $WL_{sc}$, measuring the current along the MPIL. $WL_{ss}, WL_{sg}$ represent the integral of respectively, the shear angle $\psi$ and the horizontal magnetic gradient $\nabla_h B_z$, along the MPILs strong horizontal magnetic field portions. The $WL_{sc}$ parameter corresponds to the current density $J_z$ integral along the MPILs portions of strong $B_h$. For the $WL_{ss}$ and $WL_{sc}$ parameters, only the MPILs portions where $B_h^{obs} > B_{h}^{th}$ are taken into account, while $WL_{sg}$ considers only the regions where $B_h^{pot} > B_{h}^{th}$. The two parameters $L[\Psi > 80^{\circ}]$ and $L[\psi > 80^{\circ}]$ from \citet{Leka2003a} are also tested, corresponding to the portions of the MPILs where respectively the 3D and the horizontal shear angles $\Psi$ and $\psi$ are greater than $80^{\circ}$. Finally, the unsigned flux near the MPIL is computed according to \citet{Schrijver2007}, given characteristic values around $\log R \sim 5$.  

\citet{Falconer2008} chose the following thresholds of $B_{h}^{th} = 150$~G, $\psi_{th} = 45^{\circ}$ and $\nabla_h B_z^{th} = 50$~G~Mm$^{-1}$ while we initially adopt lower thresholds for $B_{h}^{th} = 25$~G and $\nabla_h B_z^{th} = 25$~G~Mm$^{-1}$ given the small size and flux of our simulated ARs (see Section~\ref{subsec:magneto}). For the $L_{sc}$ parameter, we imposed a threshold $J_z^{th} = 12$~mA/m$^2$. It is worth to note that the choice of these thresholds can have substantial impact on the detection of eruptive-flare signatures as we will show in Section~\ref{sec:para_MPIL}, where a parametric study is proposed to estimate their influence.

\begin{landscape}
\begin{table}
\caption{Magnetic Polarity Inversion Line (MPIL) parameters tested in the present study.}             
\label{Tab:pil_para}      
\centering                          
\begin{tabular}{c c c c}        
\hline\hline                 
Description & Formula & Predictor type & Reference \\
\hline
MPIL strong-$B_h^{pot}$ length & $L_s = \int dl_{\rm{MPIL}}$, with $B_h^{pot} > B_{h}^{th}$ & $L_s$ & \citet{Falconer2008} \\[9pt]
MPIL strong-shear and $B_h^{obs}$ length & $L_{ss} = \int dl_{\rm{MPIL}}$, with $B_h > B_{h}^{th}$; $\psi > \psi_{th}$ & $L_{ss}$ & \citet{Falconer2008} \\[5pt]
MPIL strong-gradient and $B_h^{pot}$ length & $L_{sg} = \int dl_{\rm{MPIL}}$, with $B_h^{pot} > B_{h}^{th}$; $\nabla_h B_z > \nabla_h B_z^{th}$ & $L_{sg}$ & \citet{Falconer2008} \\[12pt]
MPIL strong-current and $B_h^{obs}$ length & $L_{sc} = \int dl_{\rm{MPIL}}$, with $B_h > B_{h}^{th}$; $J_z > J_z^{th}$ & $L_{sc}$ & \citet{Falconer2008} \\[12pt]
$\psi$-integral over strong-$B_h^{obs}$ MPIL & $WL_{ss} = \int \psi dl_{\rm{MPIL}}$, with $B_h > B_{h}^{th}$ & $WL_{ss}$ & \citet{Falconer2008} \\[7pt]
$\nabla_h B_z$-integral over strong-$B_h^{pot}$ MPIL & $WL_{sg} = \int \nabla_h B_z dl_{\rm{MPIL}}$, with $B_h^{pot} > B_{h}^{th}$ & $WL_{sg}$ & \citet{Falconer2008} \\[7pt]
$J_z$-integral over strong-$B_h^{obs}$ MPIL & $WL_{sc} = \int J_z dl_{\rm{MPIL}}$, with $B_h > B_{h}^{th}$ & $WL_{sc}$ & None \\[7pt]
Unsigned flux near the MPIL(s) & See Section 3 of \citet{Schrijver2007} & $R$ value & \citet{Schrijver2007} \\[7pt]
MPIL strong-shear length & $L[\Psi > 80^{\circ}] = \int dl_{\rm{MPIL}}$, with $\Psi > 80^{\circ}$ & $L[\Psi > 80^{\circ}]$ & \citet{Leka2003a} \\[7pt]
MPIL strong-3D-shear length & $L[\psi > 80^{\circ}] = \int dl_{\rm{MPIL}}$, with $\psi > 80^{\circ}$ & $L[\psi > 80^{\circ}]$ & \citet{Leka2003a} \\[7pt]
\hline                      
\end{tabular}
\end{table}
\end{landscape}

\section{Eruptive indicators evolution with $B_{mask} = 30$~G}
\label{sec:proxies}
\begin{figure}
\centering
\includegraphics[width=0.8\columnwidth]{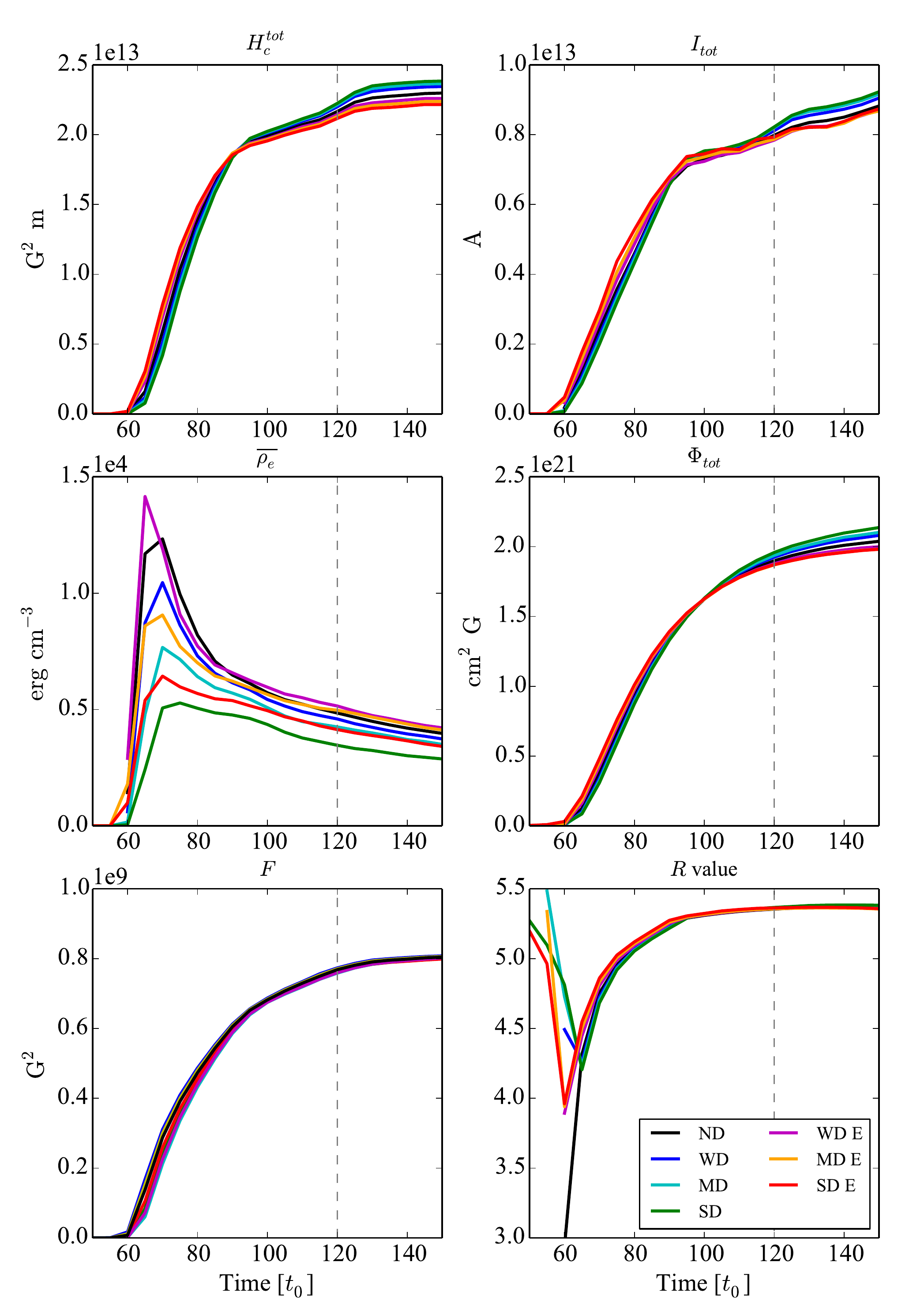}
\caption{\small Some parameters evolution for the 7 simulations, as a function of time in $t_0$ units ($t_0 = 55.5$~s). From top to bottom and left to right, the following predictor evolution are respectively displayed: the total current helicity $H_c^{\rm{tot}}$, the total current $I_{\rm{tot}}$, the mean value of the free energy $\overline{\rho_e}$, the total flux $\Phi^{\rm{tot}}$, the total Lorentz force $F$ and the Schrijver $R$ value. The warm colors indicate the evolution for the three eruptive simulations $\left\lbrace \rm{WD\ E, MD\ E, SD\ E} \right\rbrace $, respectively corresponding to the magenta, orange and red solid lines. Conversely, the evolution of the non-eruptive simulations $\left\lbrace \rm{ND, WD, MD, SD} \right\rbrace $ is displayed using the cold colors, respectively as follows: purple, deep blue, light blue and green solid lines. The eruption time, is indicated by the vertical dashed gray line. \label{fig:fig_3}}
\end{figure}

\begin{figure}
\centering
\includegraphics[width=0.8\columnwidth]{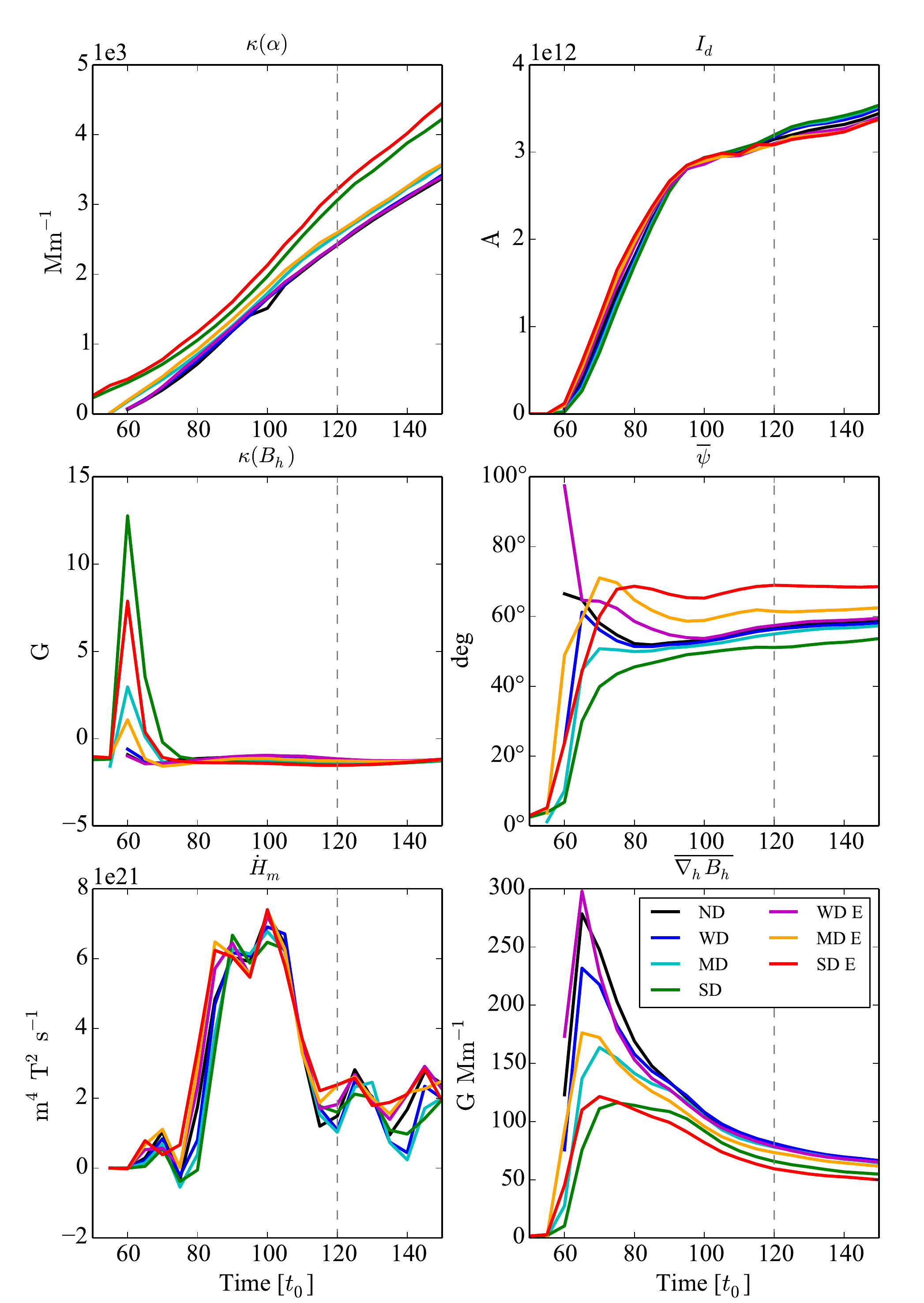}
\caption{\small Same as Figure~\ref{fig:fig_3}, but for different parameters. From top to bottom and left to right, the following indicator evolutions are respectively displayed: the kurtosis of the twist parameter $\kappa(\alpha)$, the direct current $I_{direct}$, the kurtosis of the horizontal magnetic field $\kappa(B_h)$, the mean value of the shear angle $\overline{\psi}$, the time variation of the relative magnetic helicity $dH_m/dt$ and the mean value of the vertical magnetic field $B_z$. \label{fig:fig_4}}
\end{figure}

\begin{figure}
\centering
\includegraphics[width=0.8\columnwidth]{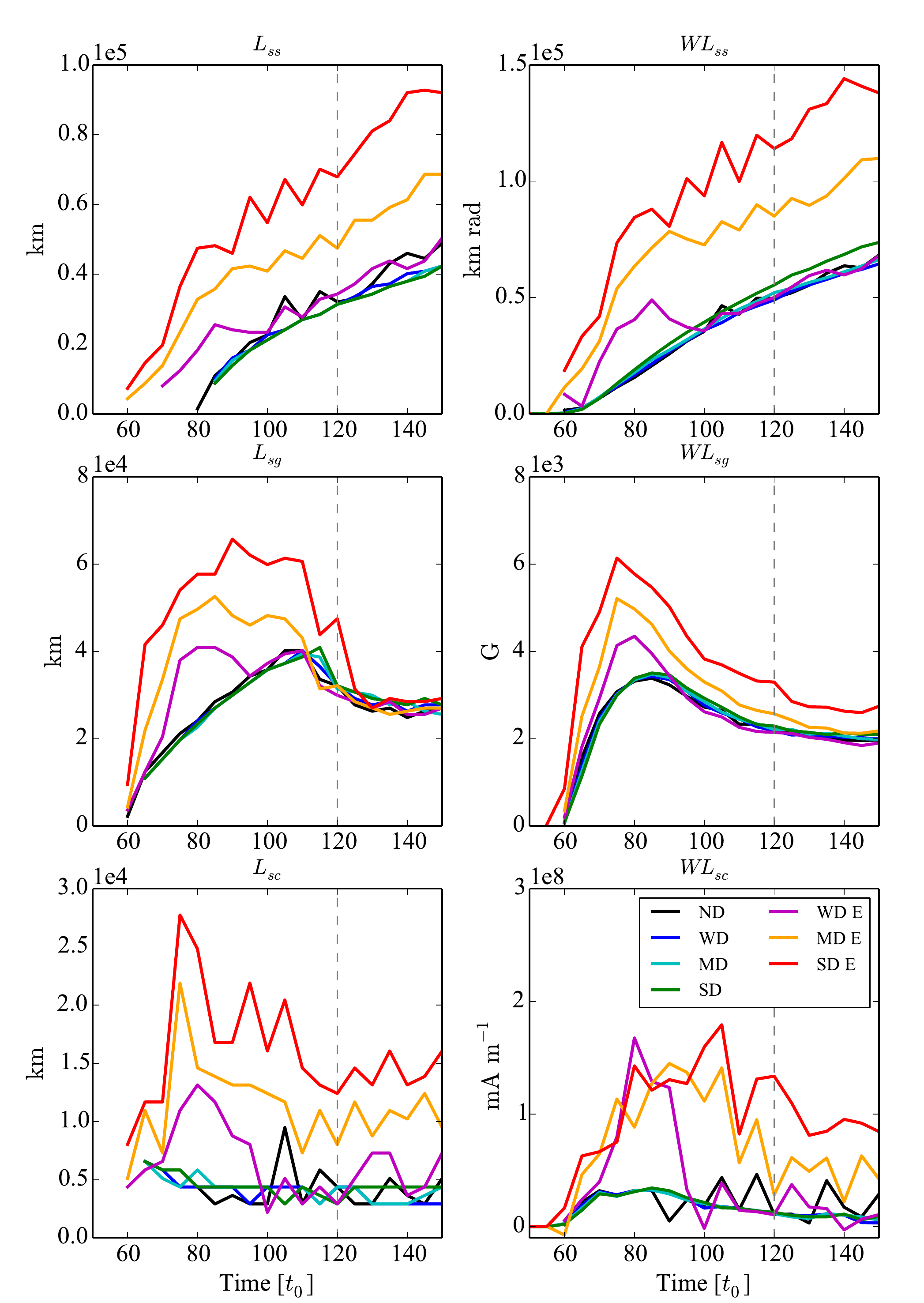}
\caption{\small Same as Figure~\ref{fig:fig_3}, but for different parameters. From top to bottom and left to right, the following indicators evolution are respectively displayed: the length of the strong shear MPIL $L_{ssa}$, the integral of shear angle along the MPIL $WL_{ss}$, the length of the strong gradient MPIL $L_{sga}$, the integral of the $B_z$ horizonal gradient along the MPIL $WL_{sg}$, the length of the strong current MPIL $L_{sc}$ and the integral of the current along the MPIL $WL_{sc}$. \label{fig:fig_5}}
\end{figure}

For each output magnetogram of each simulations, we have computed the 99 parameters described in Section~\ref{sec:parameters}. For clarity, in this section, we only present the evolution of the most representative and interesting parameters among them. The parameters shown in Figures~\ref{fig:fig_3},~\ref{fig:fig_4} and~\ref{fig:fig_5} have all been mentioned in previous studies as potential good indicators for eruptive and non-eruptive flaring activity. Even though the analysis method used in all these aforementioned works may differ from ours, i.e. using superposed epoch or forecasting methods, we used these studies as guidance to present our results and avoid quantitative comparisons. The displayed time window starts at $t = 50\ t_0$, once the flux tube has significantly emerged, the emergence starting at $t=35\ t_0$. Below $t = 50\ t_0$, none of the parameters exhibits a significant variation, and for clarity, we thus restrained the displayed time window from $t = 50\ t_0$ to $t =150\ t_0$, allowing us to analyze both prior and post eruptive flare (occurring at $t \sim 120\ t_0$) physical conditions. In the following, we do not attempt to provide physical interpretation of the full variation of the parameters, which is beyond the scope of this paper; rather we focus on the detection of significant changes in one or more parameters, according to the eruptive nature of the simulations only and occurring at a reasonable stage prior to the eruption. 

Figure~\ref{fig:fig_3} displays the evolution of some physical properties (see caption for details). The total partial current helicity $H_c^{tot}$ (top left panel), the total current $I_{tot}$ (top right panel) the free magnetic energy $\overline{\rho_e}$ (middle left panel) and the total Lorentz force $F$ (bottom left panel) are in the top-ranked eruptive indicators by \citet{Bobra2015}, whereas the total unsigned flux $\phi_{tot}$ (top right panel) and $H_c^{tot}$ have been ranked in the best predictors for 48~hours flare forecasting by~\citet{Bobra2016}. The $R$ value (bottom right panel) was empirically established as a powerful eruptive indicator by~\citet{Schrijver2007}. The parameters $H_c^{tot}$, $I_{tot}$, $\phi_{tot}$, $F$ and the $R$ value exhibit a very similar evolution, possibly due to an existing correlation between each parameter \citep{Barnes2016}. There is no evidence of physical changes prior to the eruptive flare, at about $t \sim 120\ t_0$ (vertical dashed gray line) for the eruptive simulations. 

On the other hand, the mean free magnetic energy $\overline{\rho_e}$ displays different behavior for the set of simulations, with a different peak value a $t = 70\ t_0$. The magnetic free energy is highly dependent on the region of interest selected using the $B$-mask described in Section~\ref{subsec:magneto}. This selected area increases more rapidly than the free magnetic energy per surface unit after $t = 70\ t_0$, due to the emergence process. Therefore, the mean excess magnetic energy $\overline{\rho_e}$ is observed to slowly decrease, but this is an artefact engendered by the size of the AR area. Using a constant mask as a test, the free magnetic energy is increasing as expected during the formation of a flux rope. However, the $\overline{\rho_e}$ pattern does not correlate with the flare-eruptive nature of the simulations (warm vs. cold colors solid lines), but rather with the coronal arcade magnetic field strength. Higher peak values corresponds to weaker coronal arcade, although that for a given coronal arcade magnetic field strength, the non-eruptive simulation shows systematically slightly higher values. Thus, the behavior of the $\overline{\rho_e}$ indicator is not only dependant on the eruptive nature of the simulations, but also coronal-field-strength dependant. This kind of behavior is not useful for flare forecasting, as we look for clear changes in some physical properties between stable and non-stable simulations only, in a way that some thresholds can for example be imposed. The $\overline{\rho_e}$ property changes does not allow such an approach, since a same free magnetic energy peak could correspond to both eruptive and non-eruptive ARs, with different coronal arcade magnetic strength.

Figure~\ref{fig:fig_4} presents the evolution of some magnetic field properties, such as geometry and current (see caption for details). The time evolution of the kurtosis of the twist parameter $\kappa(\alpha)$ is displayed on the top left panel. The kurtosis of the horizontal magnetic field $\kappa(B_h)$ (middle left panel) is among the more efficient predictors in the four-variable discriminant analysis of \citet{Leka2003b}, applied on flaring/non flaring ARs. The mean gradient of the horizontal magnetic field $\overline{\nabla_h B_h}$ (bottom right panel) has been found by \citet{Bobra2016} to be the best predictive variable for 24 hours flare forecasting, while the mean shear angle $\overline{\psi}$ (middle right panel) is also in the top-ten best variables for both the 24 and 48 hours predictions of the same study. We also examine the predictive capabilities of the relative magnetic helicity time variation $\dot{H}_m$ (bottom left panel) and the direct current $I_d$ (top right panel) since both have been suggested to play an important role in eruptive flare mechanisms \citep[see Section~\ref{sec:parameters} and e.g.][and references therein]{Dalmasse2015, Nindos2004}. 

As before, none of these series of parameters exhibits a clear eruptive signature. The direct current $I_{d}$ and the relative helicity variation $\dot{H}_m$ show very similar evolutions, even after the eruption at $t \sim 120\ t_0$. The kurtosis of the twist parameter $\kappa(\alpha)$ shows slightly higher values for the SD simulations, but apart from that difference, the evolution is almost the same for the whole simulation set. The $\kappa(B_h)$ presents a strong peak at $t = 60\ t_0$, with distinct values as a function of the simulations, but as for the $\overline{\rho_e}$, this behavior does not depend on the eruptive nature. Instead, the peak intensity depends on the strength of the coronal arcade, with stronger coronal magnetic field associated with stronger $\kappa(B_h)$ peak. The horizontal gradient $\overline{\nabla_h B_h}$ also presents a strong peak, at about $t \sim 66\ t_0$, whose magnitude depends once again on the overlying field strength. As the coronal field gets smaller, the horizontal gradient increases, no matter the stable/unstable nature of the simulation. Notwithstanding, the mean value of the shear angle $\overline{\psi}$ displays a slightly distinct behavior between eruptive and non-eruptive simulations. The $\overline{\psi}$ parameter exhibits somewhat greater values for the eruptive numerical experiments, notably for the MD E and SD E simulations compared to that of the stable simulations. Both the eruptive and the non-eruptive $\overline{\psi}$ remain stable once the flux tube have significantly emerged, i.e. after $t=70\ t_0$. However, there is no behavioral change before and after the eruption at $t = 120\ t_0$, a required feature for a parameter to be a reliable predictor. As the eruption can not be detected observing only the $\overline{\psi}$ parameter, the mean shear angle is not a fully discriminant parameter. 

Figure~\ref{fig:fig_5} shows the evolution for 6 of the MPIL properties extracted from the magnetogram series. All of them have been established as a powerful measure of the non-potentiality of ARs by \citet{Falconer2003, Falconer2006, Falconer2008}, and are therefore potentially good predictors for both eruptive and non-eruptive flare activity. For this class of parameters, clear eruptive flare signatures are observable, and the evolution of eruptive and non-eruptive simulations are significantly different, except perhaps for the the length of the strong-shear MPIL $L_{ss}$ (top left panel). For each of these parameters, a rise is undeniably observed during the early stage of the emergence for the three eruptive simulations, i.e. between $t=60\ t_0$ and $t \sim 95\ t_0$. During the whole emergence process, the parameter evolutions of the SD E and MD E simulations have a significantly different comportment relative to the others, while those of the WD E simulation become similar to the non-eruptive simulations once the eruption is imminent, after $t \sim 95\ t_0$. For the strong-shear length $L_{ss}$, both the SD E and MD E $L_{ss}$ lengths are significantly longer, by respectively of factor 2 and 3, than that of the stable simulations, whereas the WD E $L_{ss}$ length is barely 1.5 longer and only at the very early stage of the emergence, over a short time range. As a result, the $L_{ss}$ is not a very strong predictor, since the eruptive signature quickly disappears, at least for weak coronal magnetic field. The  parameter $WL_{sg}$ (middle right panel) shows a similar trend, with a peak at $t \sim 80\ t_0$ only about 1.2 times that of the non-eruptive simulations, making it a poorly robust predictive parameter.

However, the other parameters, $WL_{ss}$ (top right panel), $L_{sg}$ (middle left panel), $L_{sc}$ (bottom left panel) and $WL_{sc}$ (bottom right panel) are clearly more robust predictors, showing a distinct enhancement prior to the eruption, even for the WD E simulations. The $WL_{ss}$ parameter is 3 to 8.5 times greater than that of the stable simulations, while the $L_{sg}$ length is 2 to 3 times higher, but on a longer time range, between $t=75\ t_0$ and $t=95\ t_0$. The $L_{sc}$ parameter presents a sharp peak at $t = 80\ t_0$, between 2.8 and 5.4 times greater than the non-eruptive values. The $WL_{sc}$ MPIL property is the most efficient eruptive indicator, since the three eruptive simulations reach similar values at the beginning of the emergence process, being about 8 times greater than the non-eruptive $WL_{sc}$ measurements. For this latter parameter, the influence of the coronal arcade is reduced, at least during the early emergence stage. 

All these MPIL properties provide comprehensible eruptive flare signatures in these models, prior to the eruption. Apart from the $L_{ss}$ and $WL_{sg}$ parameters, for which the signature exists but is weak, the $L_{sg}$, $WL_{ss}$, $L_{sc}$ and $WL_{sc}$ parameters provide robust measurements of pre-flare eruptive conditions. As the MPIL properties are correlated with the magnetic configuration of the flux emergence, these parameters provide measurements of the magnetic complexity of the ARs. Since the eruptive flare formation and ejection are related to the quadrupolar versus bipolar nature of the emergence for this simulation set, clear eruptive flare signature can be detected. The newly defined $L_{sc}$ and $WL_{sc}$ MPIL-current properties, as well as the $WL_{ss}$ parameter appear to be the better predictor, with significant changes between eruptive and non-eruptive simulations. However, none of the other 93 parameters, related to current, magnetic field geometry or properties, provides unambiguous eruptive flare signatures, since no significant changes have been detected in their evolution.

\section{Influence of data masking}
\label{sec:data_masking}

The eruptive indicators are highly sensitive to the AR area selected to perform the computation, as seen in Section~\ref{sec:proxies}, where the flux emergence process makes the AR area increase very rapidly. \citet{Bobra2015} already highlighted that the AR parameters are highly sensitive to the data masking, and pointed out that a study to optimize such parameters is necessary. In this Section, we present, as an example, the same results as in Section~\ref{sec:proxies}, but this time using a $B_{mask} = 100$~G. This new threshold lead to a reduced AR area, and because most of the parameters are AR-area dependent, most of the results are affected. 

A variety of data masking methods can be found in previous studies. The automated system SMART for detecting ARs and their associated properties using the \textit{SOHO}/Michelson Doopler Imager \citep[MDI;][]{Scherrer1995} uses a static $B_z$ threshold of 70~G to remove the background \citep{Higgins2011}. The global features provided by the Space-Weather HMI Active Region Patches \citep[SHARPs;][]{Bobra2014} pipeline selects only the pixels for which the magnetic field strength is above the disambiguation threshold of about 150~G \citep[see also][]{Hoeksema2014}. Using the ground-based data from the University of Hawai'i Vector Magnetograph \citep{Mickey1996}, \citet{Leka2003a} used only pixels above $3\sigma$ detections. \citet{Falconer2008} imposed the $B_z$ component to be greater than 150~G, while \citet{Falconer2011} used either a threshold of 25 or 35~G, both studies using the MDI data. \citet{Algraibah2015} also used a 3$\sigma$ binary mask to exclude noisy MDI data, while~\citet{Mason2010} imposed a static threshold of 100~G.     

\begin{figure}
\centering
\includegraphics[width=0.8\columnwidth]{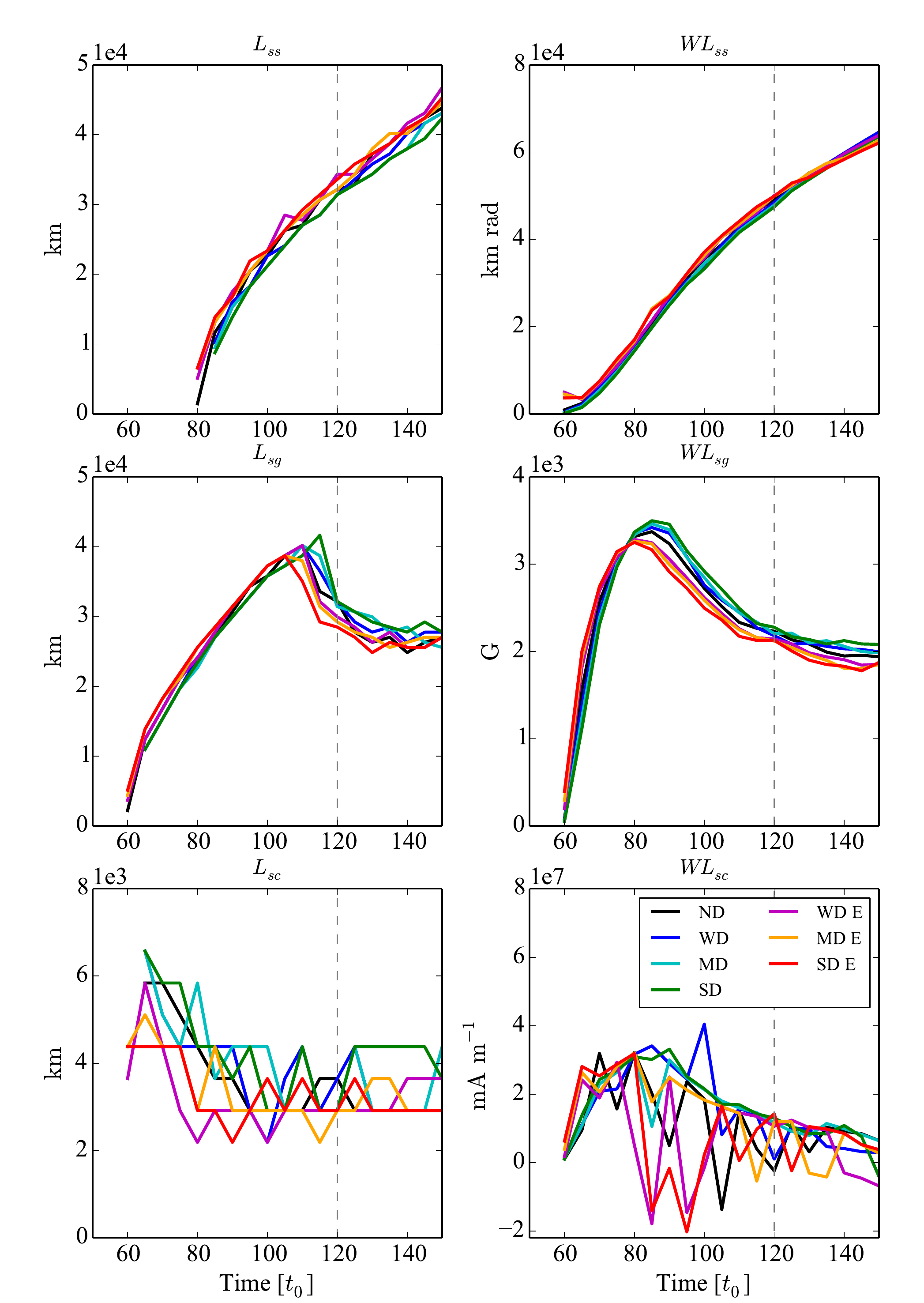}
\caption{\small Same as Figure~\ref{fig:fig_5}, but using a mask threshold $B_{mask} = 100$~G. \label{fig:fig_6}}
\end{figure}

Figure \ref{fig:fig_6} displays the same parameter evolutions as Figure~\ref{fig:fig_5}, but excluding pixels for which the total magnetic field $B$ is lower than 100~G. The seven simulations exhibit the same evolution over time, reaching similar values close to the eruption starting time. In comparison with the previous results using a lower mask threshold, the rise observed for the 6 MPIL-related parameters of the eruptive simulations is lost, and the characteristic eruptivity signature is no longer discernible. This is due to the higher threshold masking imposed to the initial data, excluding initially from the data set a fraction of the MPIL, and therefore reducing the computation domain of the MPIL properties.

\begin{figure}
\centering
\includegraphics[width=0.8\columnwidth]{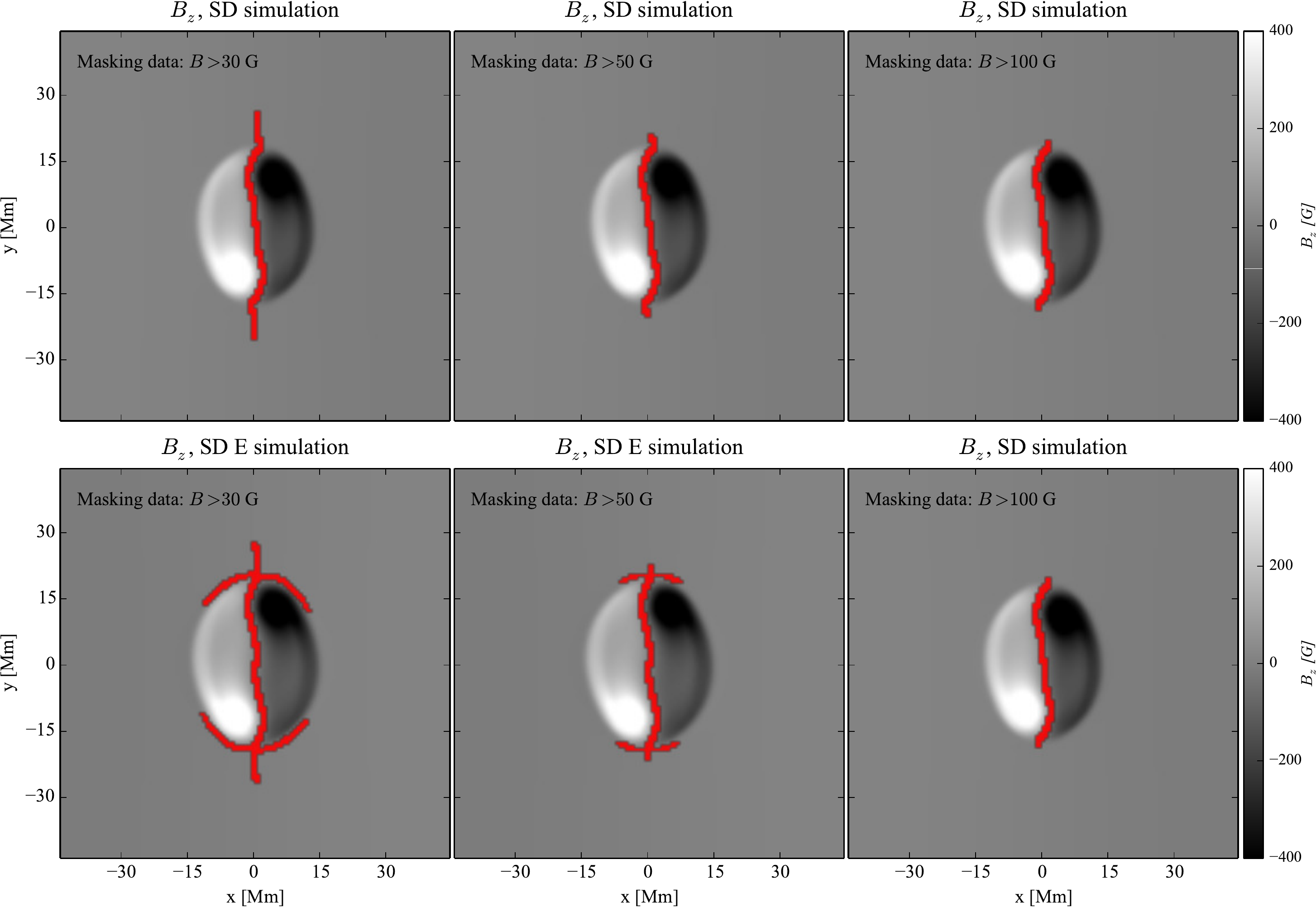}
\caption{\small Magnetograms of the SD (top row) and SD E (bottom row) simulations, including the MPIL (red lines). The MPIL width has been dilated by a factor 4 in order to increase its visibility. The masking threshold increases from left to right, with respective values of 30, 50 and 100~G. \label{fig:fig_7}}
\end{figure}

In order to visualize the data masking impact on the detection of the MPIL, Figure~\ref{fig:fig_7} displays the $B_z$ magnetograms for the SD (top row) and the SD E (bottom row) simulations, at $t=100\ t_0$. The color scale is saturated between -400 and 400~G to highlight the two polarities. Since the orientation of the arcade is opposite in SD versus SD E simulations (see Section~\ref{sec:simu} for details), the magnetic configuration is purely bipolar for the non-eruptive simulations, whereas the eruptive magnetic topology is quadrupolar. This can be clearly observed on the left panels of Figure~\ref{fig:fig_7}, where the MPIL is only located between the two polarities for the stable SD simulation (top left panel), whereas the MPIL possesses additional portions at the polarity external edges for the SD E simulation. However, as the masking threshold increases, the external MPIL is reduced, and for the case where $B_{mask} = 100$~G, both MPILs, either in the eruptive or non-eruptive simulation are almost exactly the same. Consequently, the deviation between stable and unstable simulations observed in the MPIL properties for $B_{mask} = 30$~G (see Figure~\ref{fig:fig_5}) is significantly reduced with $B_{mask} = 100$~G (see Figure~\ref{fig:fig_6}). The eruptive and non-eruptive simulations cannot be distinguished in this case, and thus the MPIL indicators become inefficient. This shows how much the initial mask threshold $B_{mask}$ should be carefully chosen, since it may be crucial to be able to detect significant eruptive signatures and therefore make reliable flare forecasting. 

\section{Impact of the noise on the detection of pre-eruptive signatures}
\label{sec:uncer}

Noise is an important issue for observation analysis. There are many different sources of error affecting the photospheric magnetic field measurements, e.g. photon noise, detector noises, spacecraft radial velocity inducing periodic systematic uncertainties~\citep{Hoeksema2014}, or uncertainties associated with the inversion of the Stokes parameters. In the present work, these different sources are not individually treated, which is beyond the scope of this paper. Instead, we simply include a random Gaussian noise to our data in order to evaluate noise impact on the detection of pre-eruptive flare signatures.

We used a Monte-Carlo scheme, randomizing the magnetograms using Gaussian perturbations, with a standard deviation of 3.5~G. The error associated with HMI magnetic field data is typically about 10~G for the line-of-sight component (V. Bommier, private communication) and can be generally one order of magnitude larger for the horizontal components \citep{Hoeksema2014}. However, our characteristic magnetic field values are about 3 to 4 times smaller than that are typically observed (see Section~\ref{subsec:scaling}). Therefore, the standard value of 10~G is not representative for our time series magnetograms. Accordingly, we adopted the characteristic noise standard deviation of 3.5~G, in order to remain consistent between our simulated observations and the relative noise scale. For this preliminary study, we assume the same error for the three components. Noise levels of 1 G and 5 G have also been tested with no sensitive differences with the results presented here. All the parameters investigated in this work are then derived from these time series of noisy magnetograms, following the same approach. First, for each simulation and for each time step, 50 different noisy magnetograms are computed by randomizing the initial one. The indicators are computed 50 times from these 50 noisy magnetogram and then averaged together. The error associated with each parameter for each time step is assumed to be the standard deviation corresponding to the 50 computed noisy indicator series, while the mean value provide the simulated magnetic field measurements.

\begin{figure}
\centering
\includegraphics[width=0.8\columnwidth]{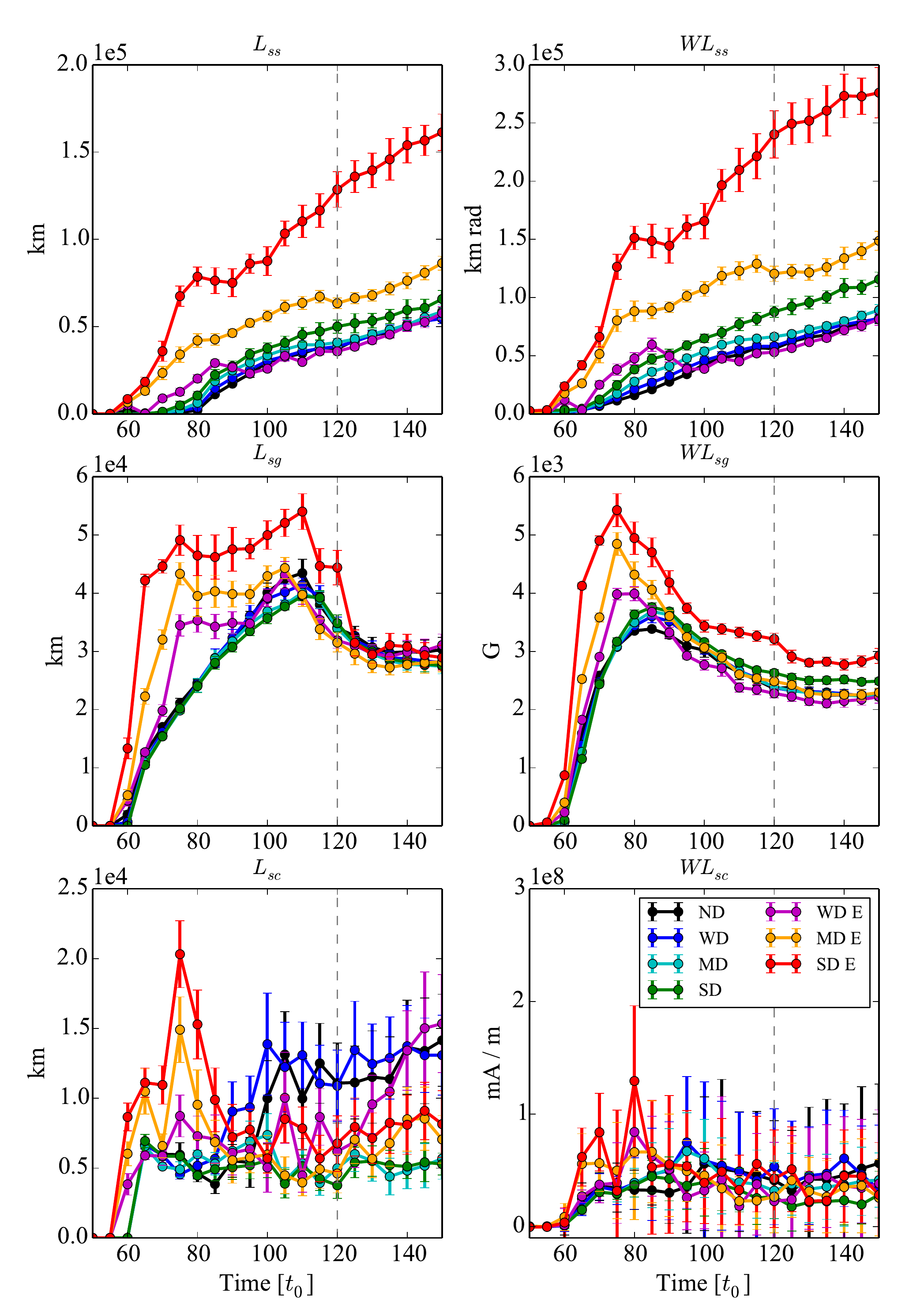}
\caption{\small Same as Figure~\ref{fig:fig_5}, but including the noise perturbation. \label{fig:fig_5_bis}}
\end{figure}

Figure~\ref{fig:fig_5_bis} displays the same parameters than Figure~\ref{fig:fig_5} but including the Monte-Carlo estimation of the measurement errors. The overall behavior remains unchanged, although some specific changes occur for the $L_{sc}$ and the $WL_{sc}$ (bottom panel) indicators. The $L_{ss}$ and the $WL_{ss}$ predictors for the MD E and SD E simulations are still significantly higher than that of the non-eruptive simulations. However, for these two parameters, in the WD E simulation, the pre-eruptive signature is lost. The sharp increase of the $L_{sg}$ and $WL_{sg}$ indicators, observed between $t=60$ and $t=100\ t_0$, is still observable, making them robust to noise perturbation. However, for the two current-related predictors $L_{sc}$ and $WL_{sc}$, the impact of noise is more important, and for instance the pre-eruptive signature provided by the $WL_{sc}$ indicator is completely dominated by the noise. For the $L_{sc}$ quantity, the trend is different, and a peak at $t \sim 75\ t_0$ is still detectable for the eruptive simulations. However, we observe a very different behavior for the WD and ND simulations compared to the clean data (see Figure~\ref{fig:fig_5}). This is due to complex effect of the noise, which generates additional pseudo-MPIL in strong-current regions.

To summarize, the presence of noise may affect the detection of the potential pre-eruptive signature, with impact depending on the parameter considered. The $L_{sg}$ and $WL_{sg}$ parameters are mostly not affected, and their pre-eruptive sharp increases are still detectable. $L_{ss}$ and $WL_{ss}$ and $L_{sc}$ are slightly affected, showing weaker pre-eruption peaks, and moderate effects for the WD E simulations. However, $WL_{sc}$ is strongly affected and the pre-eruptive signature is no longer observable.

\section{Parametric study of MPIL properties}
\label{sec:para_MPIL}

As described in Section~\ref{sec:proxies}, given our controlled-case study, only the MPIL features are able to provide a clear eruptive signature. Seven of the parameters characterizing the MPILs, namely $L_s$, $L_{ss}$, $L_{sg}$, $L_{sc}$, $WL_{ss}$, $WL_{sg}$ and $WL_{sc}$ depend on the four different thresholds $B_h^{th}$, $\psi_{th}$, $\nabla_h B_z^{th}$ and $J_z^{th}$(see Section~\ref{subsubsec:MPIL_prop} for details). These thresholds all refer to the portion of the MPILs taken into account in the computation of the various MPIL properties. As for the data masking process used to isolate the AR core area, the calculation of the eruptive indicators is highly sensitive to these values. In order to optimize the choice of these criterion and quantify their influence, we present in this section a parametric study of the MPIL properties. The threshold values explored for each parameter have been chosen based on typical values commonly used in observational data. Since our simulations correspond to small AR, we explore the parameter space using smaller and characteristic values used in previous studies.

\begin{figure}
\centering
\includegraphics[width=0.8\columnwidth]{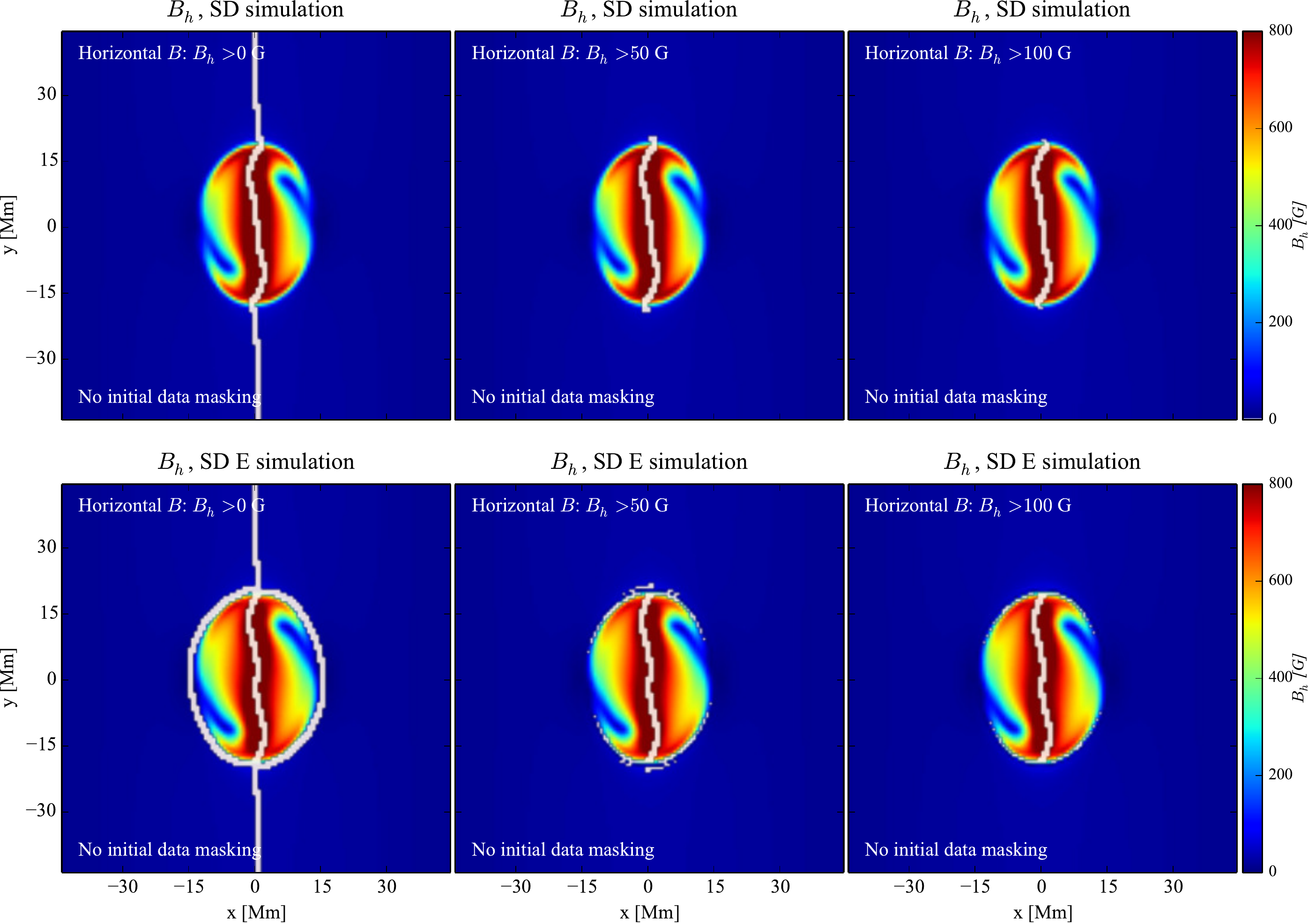}
\caption{\small Maps of the horizontal magnetic field $B_h$, for the SD (top row) and the SD E simulation (bottom row) a $t=100\ t_0$. The white line represents the portion of the MPIL for which $B_h > B_h^{th}$. The threshold increases from left to right, with values of $B_h^{th} = 0, 50$ and 100~G. No initial data masking is applied here, in order to highlight only the impact of the threshold $B_h^{th}$ on the computation of $L_{ss}$, $L_{sc}$, $WL_{ss}$ and $WL_{sc}$. \label{fig:fig_8}}
\end{figure}

\begin{figure}
\centering
\includegraphics[width=0.8\columnwidth]{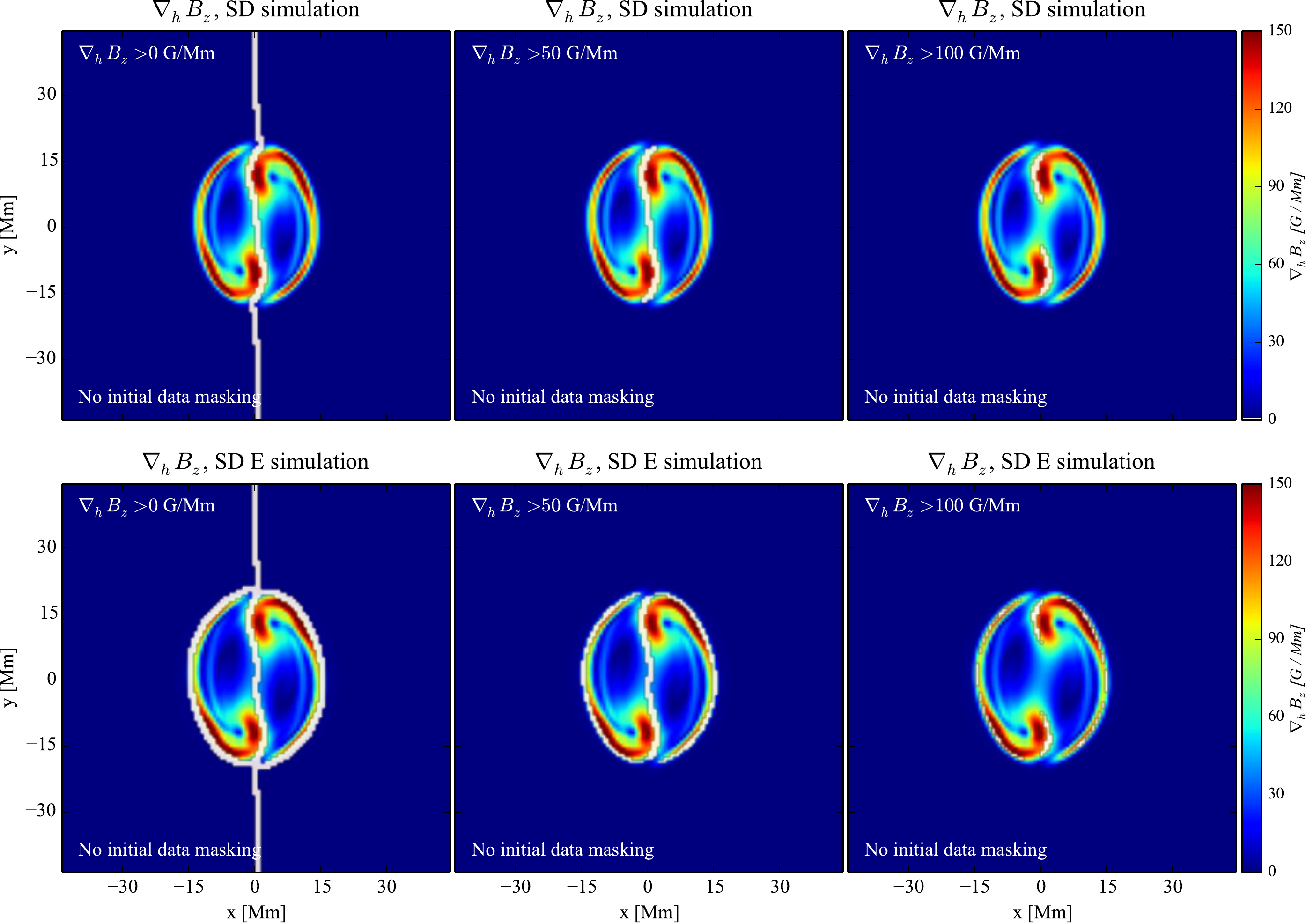}
\caption{\small Maps of the $\nabla_h B_z$ quantity, using the same layout as Figure~\ref{fig:fig_8}. The white line now represents the portion of the MPIL for which the horizontal gradient of the magnetic field $\nabla_h B_z$ is higher than $\nabla_h B_z^{th}$. From left to right the threshold $\nabla_h B_z^{th}$ increases from 0, 50 to 100 G/Mm. \label{fig:fig_9}}
\end{figure}

\begin{figure}
\centering
\includegraphics[width=0.8\columnwidth]{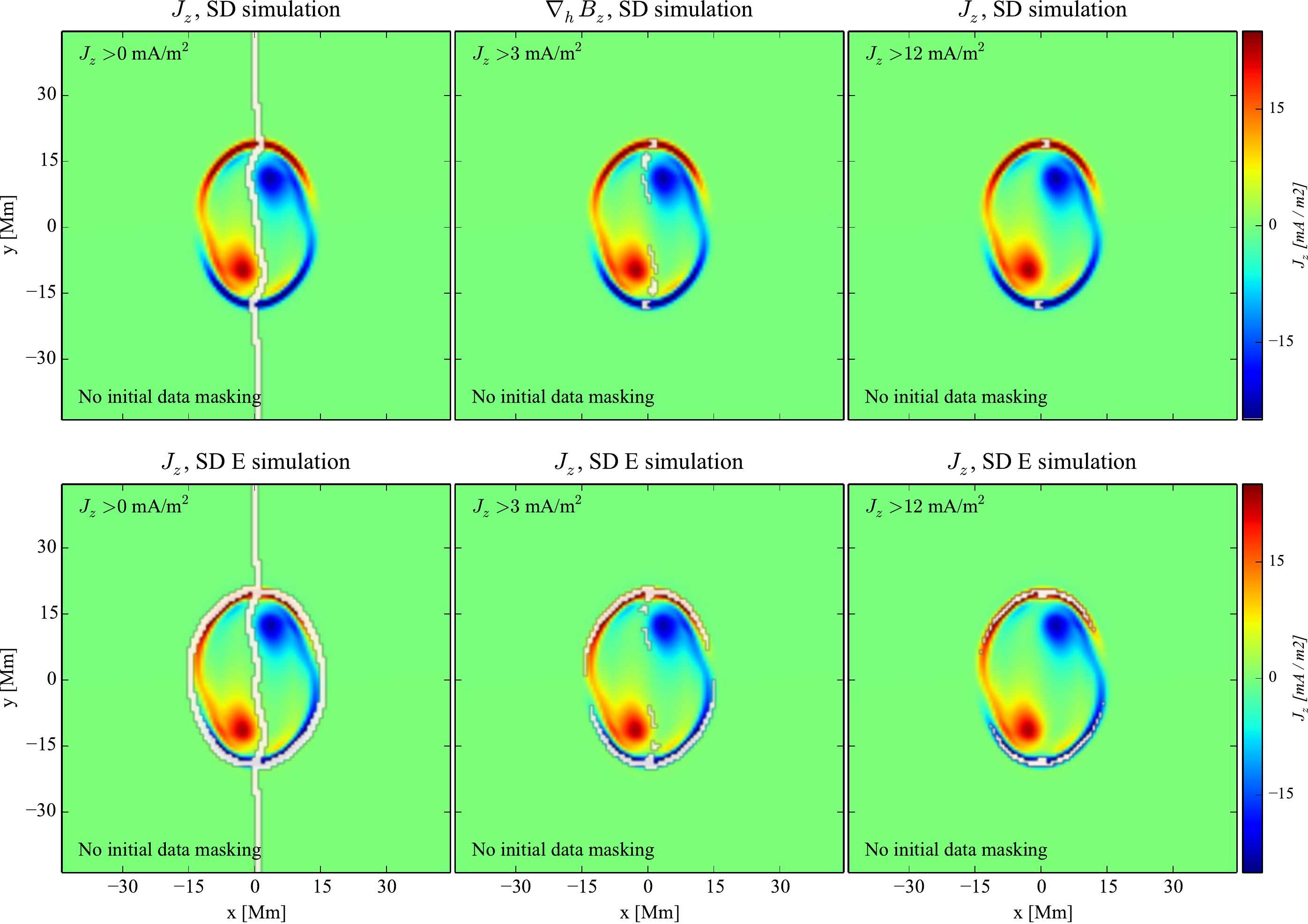}
\caption{\small Maps of the vertical current $J_z$, using the same layout as Figure~\ref{fig:fig_8}. The white line now represents the portion of the MPIL for which the current density $J_z > J_z^{th}$. From left to right, the threshold $J_z^{th}$ increases, with respective values 0, 3 and 12 mA/m$^2$. \label{fig:fig_10}}
\end{figure}

To illustrate how much the choice of the thresholds potentially affects the detection of the MPILs, Figures~\ref{fig:fig_8},~\ref{fig:fig_9} and~\ref{fig:fig_10} represent the MPIL portions (displayed as white line) detected as a function of the threshold $B_h^{th}$, $\nabla_h B_z^{th}$ and $J_z^{th}$ for the SD (top rows) and SD E (bottom rows) simulations, in a similar fashion than Figure~\ref{fig:fig_7}. The length of the MPIL (white line) depends on these different parameters. Figure~\ref{fig:fig_8} displays $B_h$ maps on the background, and the $B_h^{th}$ parameters varies from left to right, while Figure~\ref{fig:fig_9} represents $\nabla_h B_z$ maps as a function a the $\nabla_h B_z^{th}$ threshold. Figure~\ref{fig:fig_10} displays $J_z$ maps and the $J_z^{th}$ threshold changes from 0 (left) to 12~mA/m$^2$ (right). In this section, the objective is to start to investigate the effects of each thresholding parameter on the MPIL parameters that have been demonstrated to be efficient eruptive predictors, under the condition that the initial mask thresholding was low enough (see Section~\ref{sec:proxies}). In this section, we do not initially mask the data (see Section~\ref{subsec:magneto}, i.e. $B_{mask} = 0$~G) as we focus on threshold impact. Obviously, it is worth noting that the conjugate effects of high mask threshold (see Sections~\ref{subsec:magneto} and~\ref{sec:data_masking}) and individual parameter thresholdings are even stronger on the detection of the eruptive signature and therefore highly worsen the results.

The $B_h^{th}$ threshold has a strong effect on the detection of the whole MPIL: as seen in Figure~\ref{fig:fig_8}, the quadrupolar configuration, corresponding to outer MPIL on the edge of the polarities (see the bottom left panel), is no longer detected as $B_h^{th}$ is greater than 50~G, therefore corresponding to similar MPIL (white line on right panels). As such, any MPIL properties computed using such a threshold will not provide reliable eruptivity signature. The $\nabla_h B_z^{th}$ threshold is less restrictive, and a small fraction of the MPIL outer portions (white line on the bottom right panel) is still detected for the SD E simulations. However, it is worth noting that the MPIL represented in Figure~\ref{fig:fig_9} is dilated to increase the visibility; actually the MPIL is thinner and only a few additional pixels are detected. On the other hand, the $J_z^{th}$ thresholding does not affect that much the larger measurements of eruptive MPIL: as seen in the bottom right panel of Figure~\ref{fig:fig_10}, the external portions of the MPIL are still observed, even using a threshold of 12~mA/m$^2$. From these series of maps as a preliminary coarse evaluation of thresholding effects on the MPIL properties, we can conclude that the current thresholding $J_z^{th}$ is the most flexible parameter, since the quadrupolar configuration of the eruptive simulations can still be detected even when increasing threshold. 

\begin{figure}
\centering
\includegraphics[width=0.8\columnwidth]{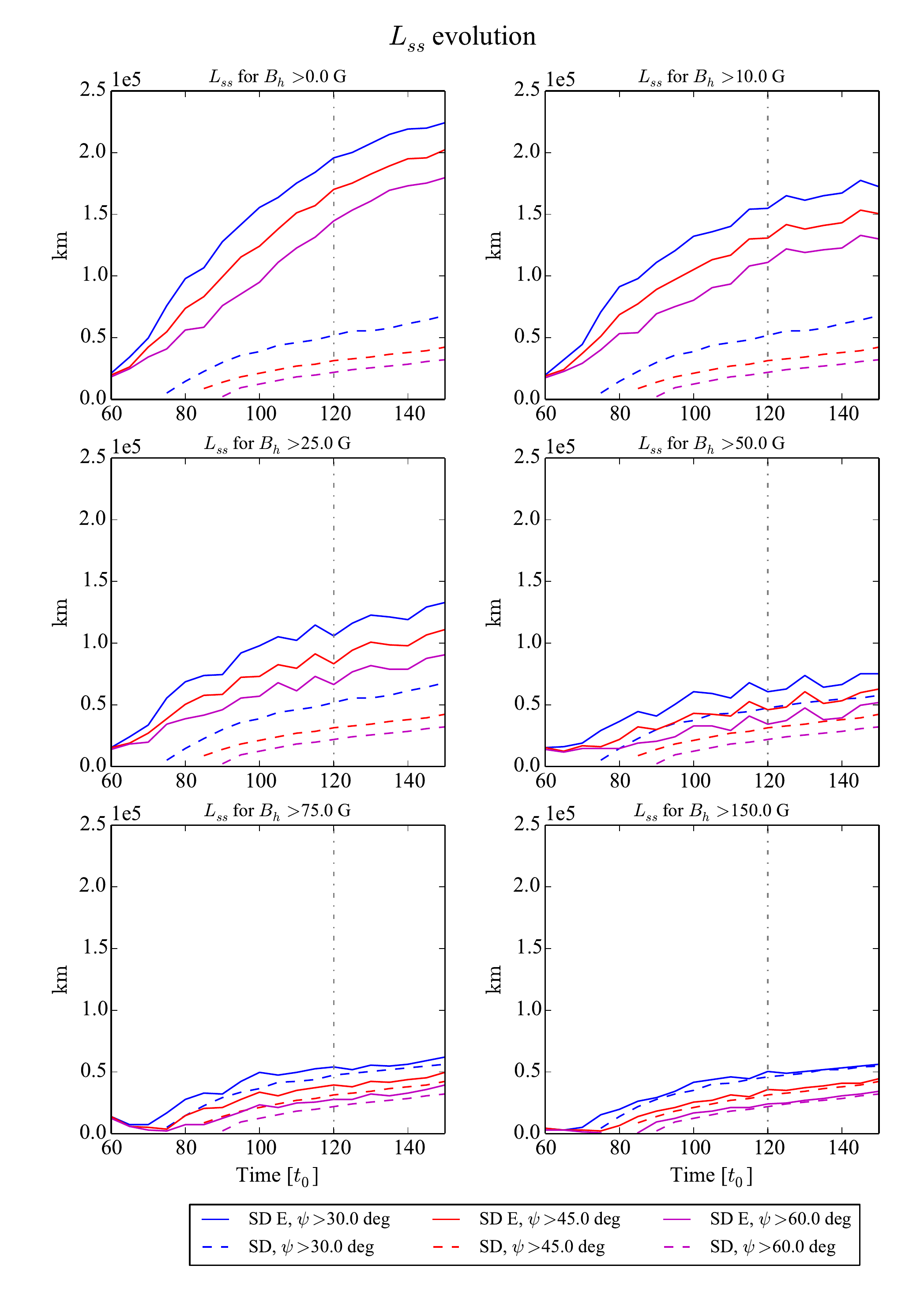}
\caption{\small Parametric evolution of the total length of the strong shear MPIL $L_{ss}$, as a function of the two thresholds $B_h^{th}$ and $\psi_{th}$. The $B_h^{th}$ varies for each figure, with respective values of 0, 10, 25, 50, 75 and 150~G, from top to bottom and left to right. For each panel, the $L_{ss}$ quantity is represented for the strong arcade eruptive (SD E; solid lines) and strong arcade (SD; dashed lines) simulations. The threshold $\psi_{th}$ is changed for each curve, using the values of 30 (blue lines), 45 (red lines) and 60$^{\circ}$ (magenta lines). The eruption time is denoted by the vertical gray dot dashed line. \label{fig:fig_11}}
\end{figure}

To further investigate the influence of the physical thresholds on the MPIL predictors, we compute the variations of the $L_{ss}$, $L_{sg}$, $L_{sc}$, $WL_{ss}$, $WL_{sg}$ and $WL_{sc}$ indicators as a function of their associated threshold. The three first quantities depend on two different thresholds, making therefore their evolution even more sensitive to the chosen cutting values. Figure~\ref{fig:fig_11} displays the $L_{ss}$ variations for both eruptive SD E (solid lines) and non-eruptive SD (dashed lines) simulations, as a function of the $B_h^{th}$ and $\psi_{th}$ thresholds. The $B_h^{th}$ is increasing from panel to panel, while the $\psi_{th}$ is fixed for a given color line. If the $B_h^{th}$ is below 25~G (top and middle left panels), we can still observe longer $L_{ss}$ lengths for the SD E simulation, whatever the $\psi_{th}$ threshold. In these three cases, the SD E $L_{ss}$ parameter is respectively about 4, 3 and 2 times longer than that of the SD simulation, a parameter difference still measurable. However, for higher $B_h^{th}$, this difference is reduced and discriminate between eruptive and non-eruptive simulations becomes difficult. On the other hand, the influence of the $\psi_{th}$ threshold is weak and whatever the chosen value, SD E and SD simulations can still be distinguished.    

\begin{figure}
\centering
\includegraphics[width=0.8\columnwidth]{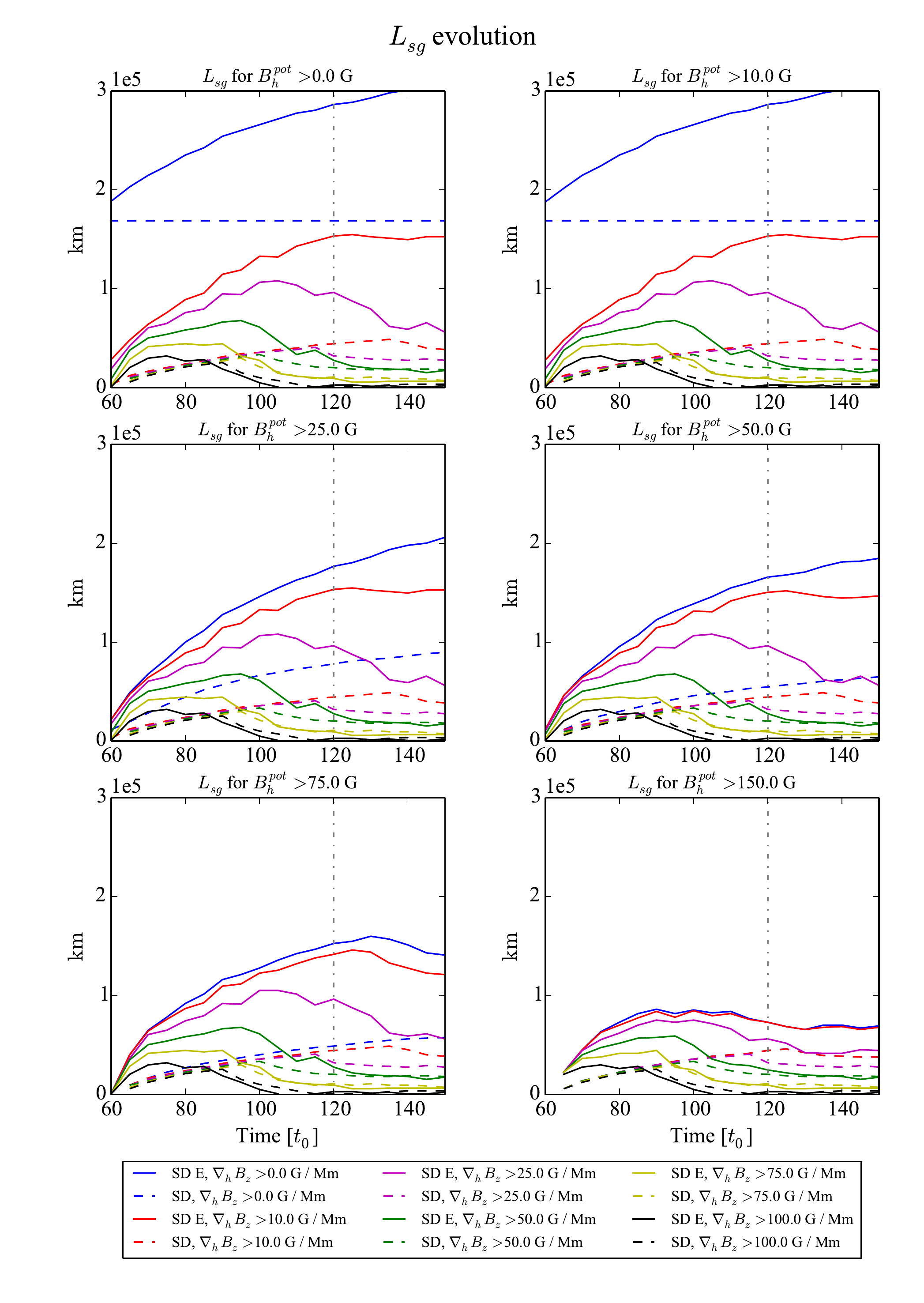}
\caption{\small Same as Figure~\ref{fig:fig_11} but for the parameter $L_{sg}$, as a function of the two thresholds $B_h^{th}$ and $\nabla_h B_z^{th}$. The $B_h^{th}$ varies in the same way as $B_h^{th}$ for Figure~\ref{fig:fig_11}. The threshold $\nabla_h B_z^{th}$ is changed for each curve, using the values of 0 (blue lines), 10 (red lines), 25 (magenta lines), 50 (green lines), 75 (yellow lines) and 100~G/Mm (black lines). \label{fig:fig_12}}
\end{figure}

Figure~\ref{fig:fig_12} is the same as Figure~\ref{fig:fig_11}, but exploring the variations of the $L_{sg}$ predictor as a function of $B_h^{th}$ and $\nabla_h B_z^{th}$ thresholds. As before, each panel corresponds to a given $B_h^{th}$ threshold, and each color line corresponds to a given $\nabla_h B_z^{th}$. For this parameter, the influence of the $B_h^{th}$ threshold is lower, while the impact of the $\nabla_h B_z^{th}$ is significantly stronger. For example, for the given $\nabla_h B_z^{th} = 10$~G/Mm (red lines on each panel), whatever the $B_h^{th}$ imposed, the SD E $L_{sg}$ length is still longer than that of the SD simulation. However, as the $\nabla_h B_z^{th}$ increases, the divergence between the SD E and SD curves rapidly disappears, making the eruptive signature undetectable. For the two top panels, it is worth noting that the length $L_{sg}$ (dashed blue line) of the SD simulation remains constant over the whole numerical experiment. This is due to the low $B_h^{th}$ imposed, below the photospheric arcade magnetic field mean value, conjugated to the $\nabla_h B_z^{th} = 0$~G/Mm threshold, allowing therefore to measure the whole MPIL on the entire numerical domain. Since SD is a non-eruptive simulation, the magnetic field configuration is bipolar, and the MPIL is almost kept constant. 

\begin{figure}
\centering
\includegraphics[width=0.8\columnwidth]{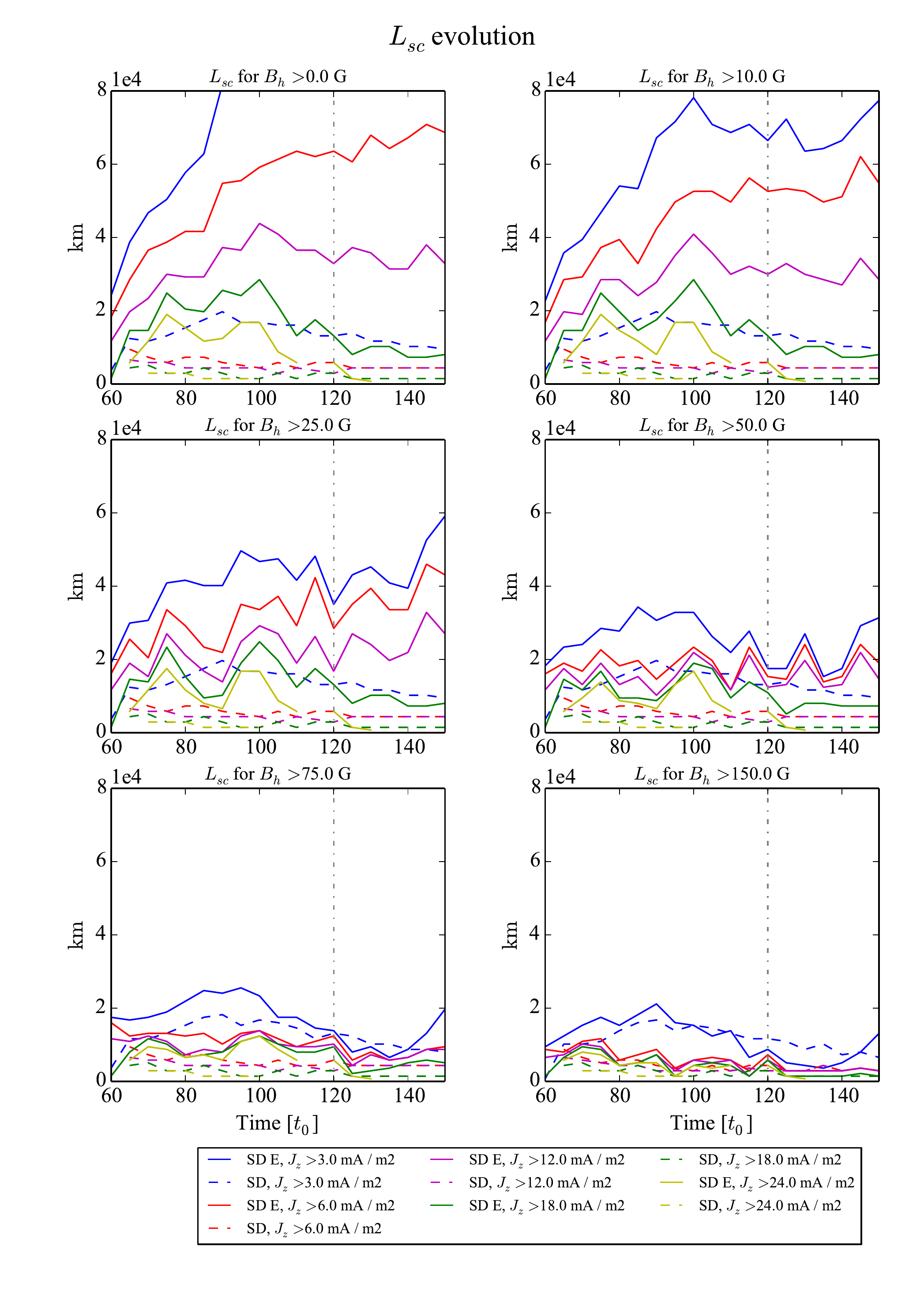}
\caption{\small Same as Figure~\ref{fig:fig_11} but for the parameter $L_{sc}$, as a function of the two thresholds $B_h^{th}$ and $J_z^{th}$. The threshold $J_z^{th}$ is changed for each curve, using the values of 3 (blue lines), 6 (red lines), 12 (magenta lines), 18 (green lines) and 24~mA/m$^2$ (yellow lines). \label{fig:fig_13}}
\end{figure}

Figure~\ref{fig:fig_13} shows in the same way as Figures~\ref{fig:fig_12} and Figure~\ref{fig:fig_11} the variations of the newly introduced $L_{sc}$ parameter, as a function of its two associated thresholds $B_h^{th}$ and $J_z^{th}$. For this MPIL property, the influence of both threshold is important. As for the $L_{ss}$ parameter, detecting the eruptive signature, i.e. a longer $L_{sc}$ for the eruptive simulations, becomes harder as $B_h^{th}$ exceeds 25~G. For $B_h^{th} = 50$~G, a longer $L_{sc}$ for the SD E simulation is however still detectable, but the difference between eruptive and non-eruptive simulations is less obvious. For higher $B_h^{th}$ (bottom panels), discriminating the simulations is impossible, whatever the current $J_z^{th}$. On the other hand, increasing the $J_z^{th}$ has also a strong effect on the detection of the eruptive signature, even though small differences between eruptive and non-eruptive simulations persist. For instance, if $B_h^{th} = 10$~G (top right panel) and $J_z^{th} = 3$~mA/m$^2$ (blue lines), $L_{sc}$ is almost 4 times longer before the eruption for the SD E simulation than that of the SD. If we increase $J_z^{th}$ to $24$~mA/m$^2$ (yellow lines), the $L_{sc}$ SD E to SD ratio decreases to 1.7.     

From these three plot analyses, general dependence trends can be deduced. The $L_{ss}$ parameter is strongly dependent on the $B_h^{th}$ threshold, and for $B_h^{th}$ greater than 50~G, the eruptivity signature is no longer measurable. However, the influence of the shear-angle threshold $\psi_{th}$ is rather weak, and whatever the threshold assumed, $L_{ss}$ parameters still allow us to discriminate between the SD E and SD simulations. Conversely, the $L_{sg}$ threshold is weakly dependant on the $B_h^{th}$ threshold, while the gradient $\nabla_h B_z^{th}$ threshold has a strong impact on the detection of eruptive ARs. For $\nabla_h B_z^{th} > 50$~G/Mm, the distinction between both ARs types is hardly observable. Finally, the $L_{sc}$ quantity is impacted by the choice of both its associated threshold, even if small remnants of eruptive signature are still detectable for high thresholding values.

\begin{figure}
\begin{tabular}{cc}
\includegraphics[scale=0.38]{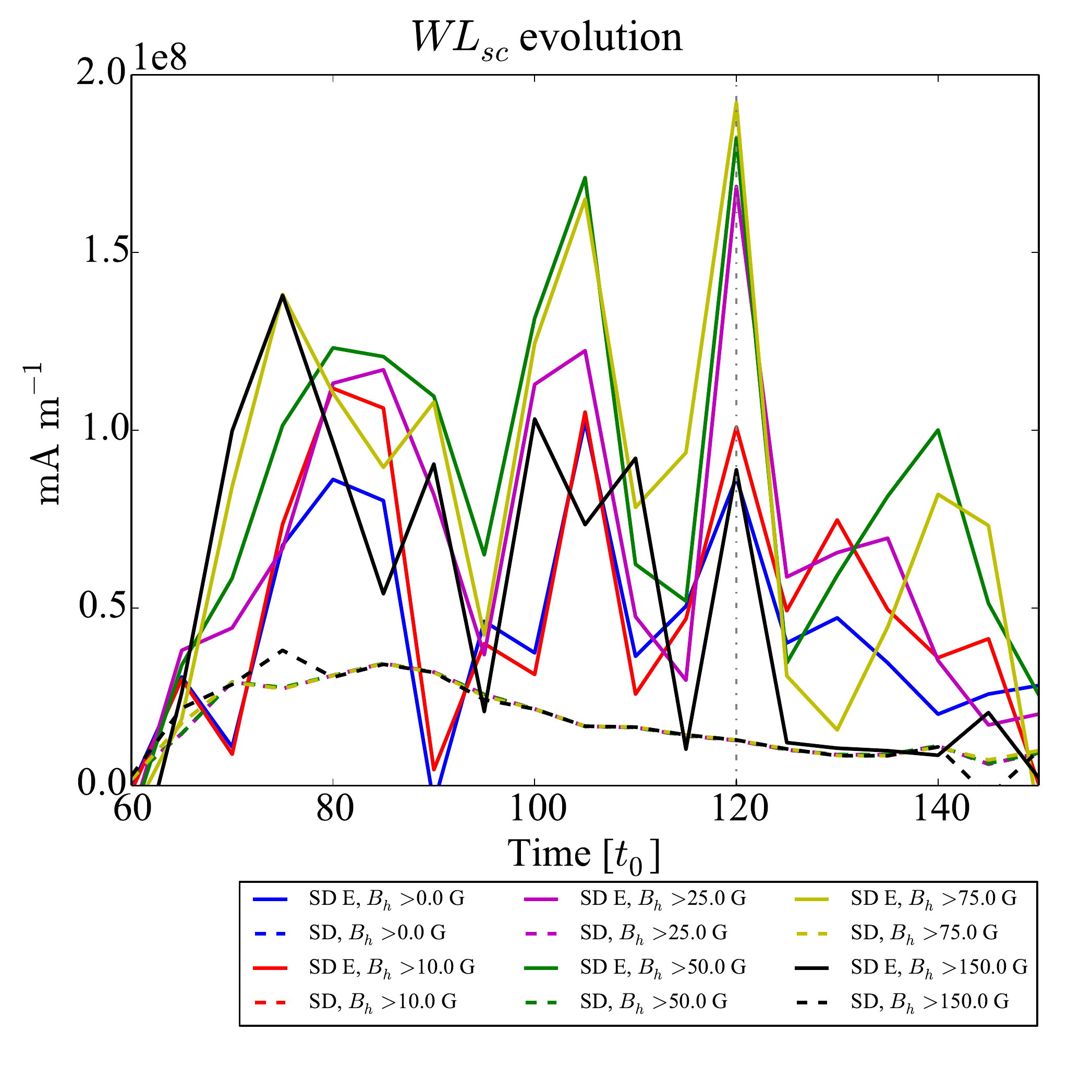} &
\includegraphics[scale=0.38]{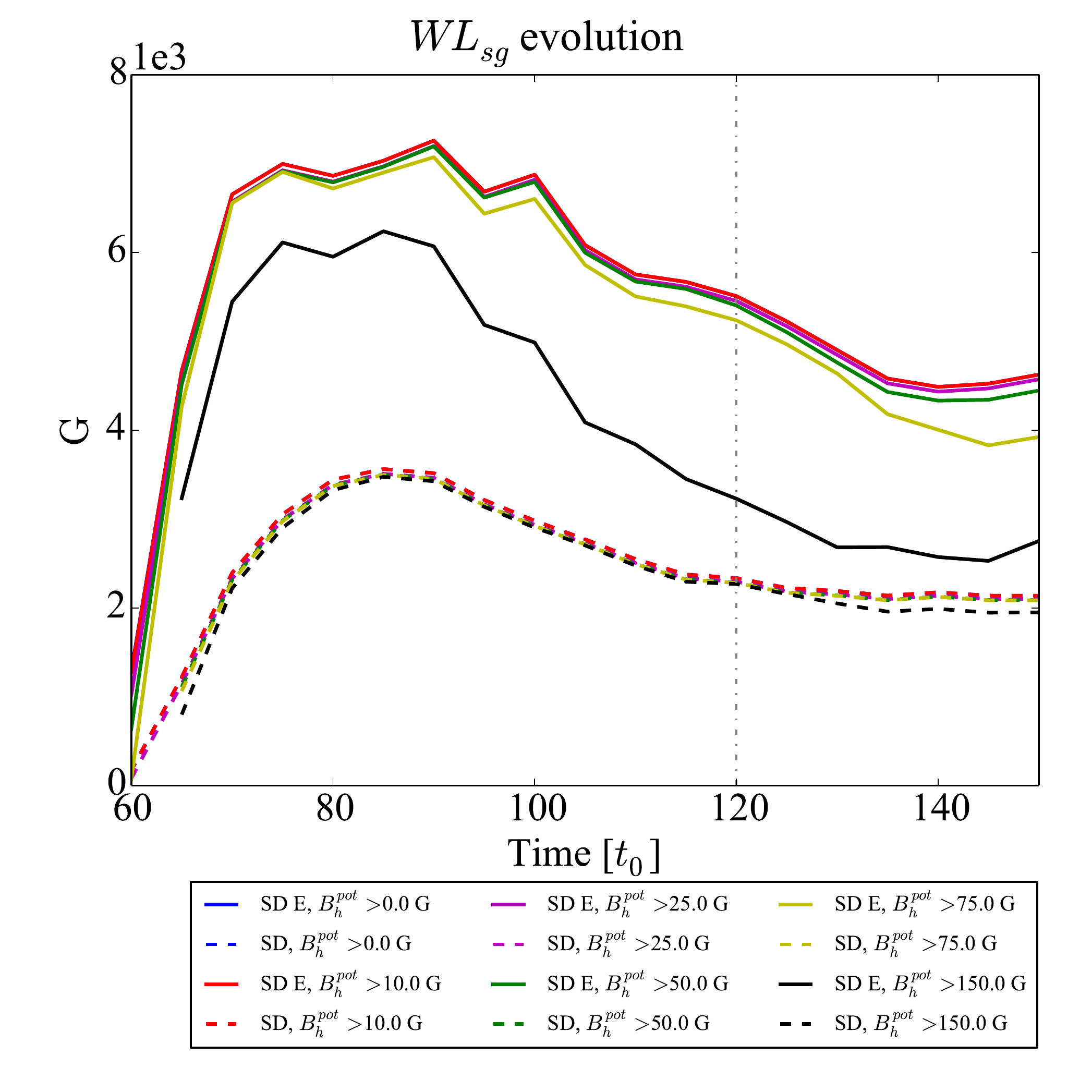} \\
\includegraphics[scale=0.45]{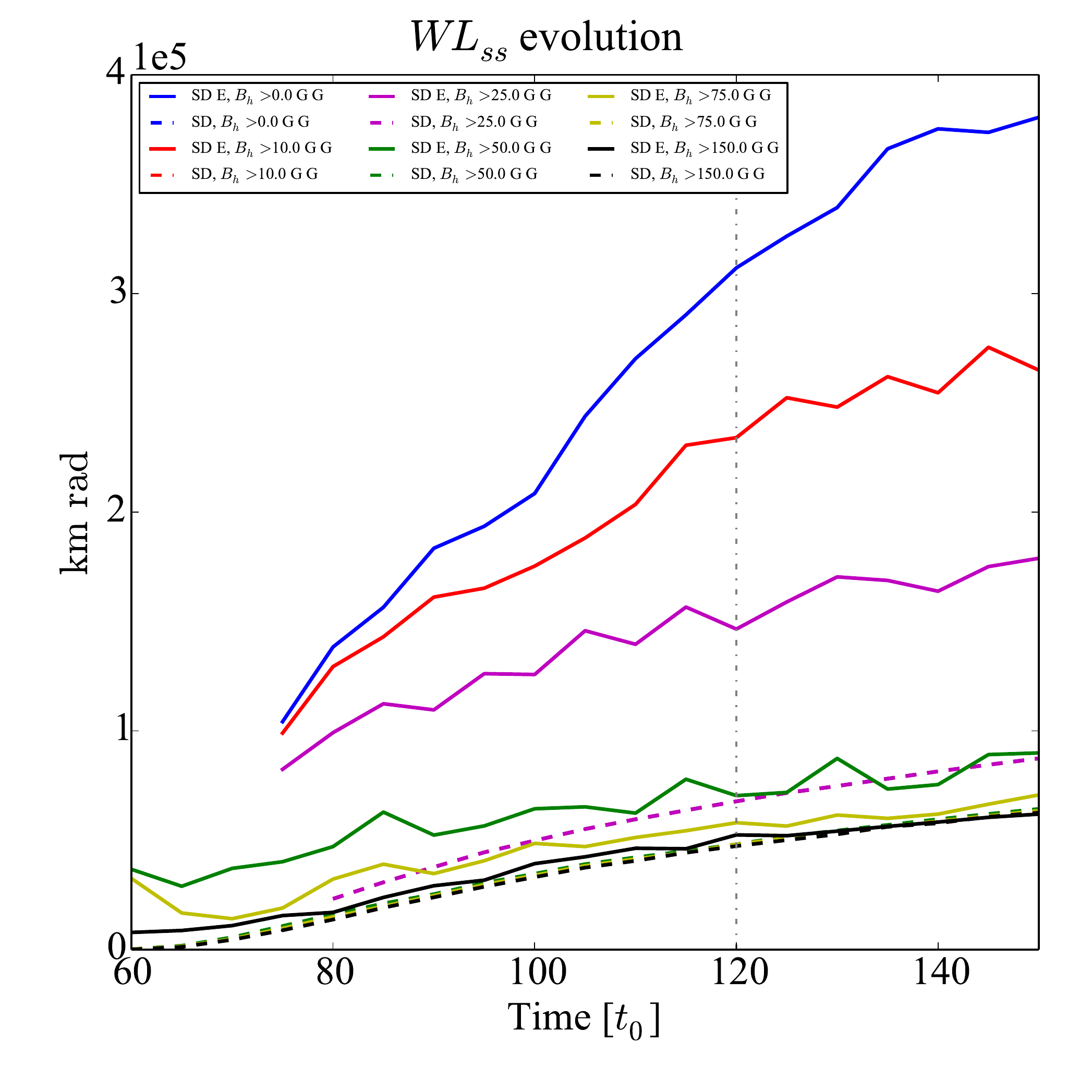} & \\
\end{tabular}
\caption{\small Parametric evolution of the $WL_{ss}$ (left panel), $WL_{sg}$ (middle panel) and $WL_{sc}$ (right panel) parameters, as a function of the threshold $B_h^{th}$. Solid lines represents the parameter evolutions for the SD E simulation, while dashed lines are computed for the SD stable simulation. The $B_h^{th}$ varies from 0 (blue lines) to 150~G (black lines), with the following intermediate values 10 (red lines), 25 (magenta lines), 50 (green lines) and 75~G (yellow lines). \label{fig:fig_14}}
\end{figure}

The $WL_{ss}$, $WL_{sg}$ and $WL_{sc}$ parameters only depend on the $B_h^{th}$ threshold (see Table~\ref{Tab:pil_para}). Figure~\ref{fig:fig_14} displays the variations of these indicators for both the SD E (solid lines) and the SD (dashed lines) simulations, varying the $B_h^{th}$ threshold, in order to optimize the choice of this parameter. The left panel of Figure~\ref{fig:fig_14} displays the $WL_{ss}$ parameter evolutions, and as the $B_h^{th}$ threshold increases, the $WL_{ss}$ parameter decreases for the SD E simulation while it remains stable for that of the SD non-eruptive. Using a low excluding value allows us to detect the external portion of the MPIL (blue, red and magenta lines) and therefore to discriminate between the eruptive and non-eruptive simulations, while thresholds larger than 50~G (green, yellow and black lines), remove this eruptive signature, characterized by a larger $WL_{ss}$ for eruptive ARs. 

The middle and right panels of Figure~\ref{fig:fig_14} present respectively the $WL_{sg}$ and $WL_{sc}$ parametric analysis. For these two parameters, the thresholding impact is weak, and even using high threshold allow to distinguish between eruptive and non-eruptive simulations. The $WL_{sg}$ property (see the middle panel of Figure~\ref{fig:fig_14}) is about 2-2.5 times higher for the SD E simulation than that of the SD, whatever the adopted threshold $B_h^{th}$ imposed on $B_h^{pot}$. The newly defined $WL_{sc}$ parameter show sharp variations for the eruptive simulation, increasing quickly by a factor about 4, whereas the non-eruptive evolution remains smooth over the whole emergence process. These sharp evolutions are observed whatever the $B_h$ threshold imposed on the horizontal observed magnetic field, demonstrating that the noise-free $WL_{sc}$ is a robust predictor for detecting flare producing ARs.

From these parametric studies of the MPIL properties $WL_{ss}$, $WL_{gs}$ and $WL_{sc}$, clear conclusions can be drawn. The $WL_{ss}$ predictor is highly dependent on the threshold process, and the eruptive signature, corresponding to higher $WL_{ss}$ in eruptive ARs is rapidly lost as $B_h^{th}$ increases. Consequently, if the $WL_{ss}$ is able to detect imminent eruption in an AR under certain conditions (i.e. low masking threshold $B_{mask}$, see Section~\ref{sec:data_masking}), this may not be the most reliable eruptive indicator to be used for flare forecasting, unless different thresholds are simultaneously tested. On the other hand, both noise-free $WL_{sg}$ and $WL_{sc}$ parameters appears as very robust eruptive predictors since the choice of the threshold $B_h^{th}$ does not affect the detection of the eruptive signature. 

\section{Summary and Conclusions} 
\label{sec:summary}  

The predictability of magnetic properties has been barely estimated using numerical simulations (see Section~\ref{sec:intro}). In this work, we have studied the reliability of some eruptive flare indicators using the 3D parametric MHD simulations of flux emergence in~\citet{Leake2013, Leake2014}, leading to the self-consistent formation of either stable or unstable flux ropes. A set of 4 stable and 3 eruptive simulations have been used, corresponding to the partial emergence of a twisted magnetic flux tube into a stratified atmosphere, where a coronal overlying field is present. The eruption is triggered by a combination of external magnetic reconnection between the dipole and the overlying magnetic field, followed by internal reconnection allowing the vertical expansion and thus the ejection of the flux rope. The behavior of the emergence has been investigated through a range of initial coronal arcade strength and orientation. Depending on the coronal arcade orientation, the magnetic configuration is either dipolar or quadrupolar, leading to respectively stable or unstable coronal flux rope (see Section~\ref{subsec:mhd}).

Parameters have been examined in order to detect whether some physical changes specific to eruptive-flaring ARs could be detected and measured, thus establishing a certain relation to flare occurrence. We rigorously analyzed the simulations as observations, including data masking and magnetic flux transport velocity derivation from the DAVE4VM code~\citep{Schuck2008}, using time series of 2D-plane magnetograms extracted from the 3D numerical datasets. A list of 99 parameters has been tested, including indicators currently used for both eruptive and non-eruptive flare operational forecasting as well as new quantities, such as helicity or current-weighted neutral line lengths $L_{sc}$ and $WL_{sc}$.   
 
From the 99 predictors list used in this work, only the parameters relying on MPIL properties (see Table~\ref{Tab:pil_para}) demonstrated predicting eruptive flare capabilities. The other parameters showed a very similar evolution for both stable and unstable simulations, and no eruptive signature was detectable (see Section~\ref{sec:proxies}). The strong-shear MPIL length $L_{ss}$, the strong-gradient MPIL length $L_{sg}$ and the strong-current MPIL length $L_{sc}$ are significantly longer for the unstable simulations due to the quadrupolar magnetic configuration generating additional external MPIL (see e.g. Figure~\ref{fig:fig_7}). The length increase rate depends on the coronal magnetic field strength: stronger coronal dipole are associated to longer property-weighted MPIL lengths. However, a strong rise of these three lengths is systematically observed during the early stage of the emergence, whatever the arcade field strength. The $WL_{ss}$, $WL_{sg}$ and the $WL_{sc}$ parameters present very similar trends, making also them promising eruptivity predictors.  

The detection of eruptive flare signatures, i.e. higher MPIL properties for eruptive ARs, are dependent on the initial masking process. Indeed, to isolate the ARs from the background magnetic field, a thresholding mask is usually applied when observations are analyzed. Notwithstanding, we demonstrated that the choice of the initial masking threshold is crucial for detecting MPIL properties variations. The same analysis has been presented for both $B_{mask} = 30$ and 100~G (see Section~\ref{sec:data_masking}), showing that the eruptive flare signatures disappear using the higher mask threshold. Noise in the measurements can also scramble the eruptivity signatures: through a Monte-Carlo scheme, we estimated the noise influence by including random perturbation to our time series magnetograms. Our results show that the four $L_{ss}$, $L_{sg}$, $WL_{ss}$ and $WL_{sg}$ parameters are not strongly impacted by noise and their associated peaks prior to the eruption is still detectable. However, the $L_{sc}$ and $WL_{sc}$ are more strongly impacted: the increase of $L_{sc}$ is weaker, although still measurable, but the $WL_{sc}$ parameter no longer allows to discriminate between eruptive and stable simulations.      

In addition, the MPIL properties, associated with valuable eruptive flare predictabilities, depend not only on the initial masking process and the noise, but also on additional thresholds imposed on physical properties, such as current, magnetic field or gradients. We also investigated the impact of these physical threshold through a parametric study measuring the detectability of eruptive signature as a function of the thresholding process (see Section~\ref{sec:para_MPIL}). Results show that the three MPIL lengths $L_{ss}$, $L_{sg}$ and $L_{sc}$ and the $WL_{ss}$ parameter are strongly sensitive to the choice of the physical threshold, whereas $WL_{sg}$ and $WL_{sc}$ are robust relative to threshold changes.

Our study is limited in term of AR size and complexity that we can not explore given the computational cost of such simulations. Given the small scale of our MHD simulations, we do not catch the same distribution of magnetic field strengths as observations. Because of the small flux (in the $10^{21}$ Mx range), and the absence of interaction between granulation and magnetic field, the intermediate strength fields are not caught. Hence, by applying the $B$-mask, which is done at these intermediate strengths, some features from the simulations may be lost. Besides, eruptive flare indicators have only been tested for a given AR flux and size, corresponding to the smallest observed flaring ARs class (see Section~\ref{subsec:scaling} for details). Still, this type of analysis is yet relevant to connect flare physical models and observations, and provide a comprehensive perspective of what can be done using actual observations. Future work will be extended to AR simulations with different sizes and fluxes.   

We therefore conclude that from our 99 predictors list, $WL_{sg}$ and $L_{sg}$ are the best eruptive flare indicators for these model tests. The other parameters relying on MPIL properties tested in this study, namely $L_{ss}$ and $L_{sc}$, $WL_{sc}$ and $WL_{ss}$ should be tested using various physical thresholds. The current-related parameters seems to be more sensitive to noise, even if a more detailed analysis of uncertainty sources is needed. Apart from the physical thresholds, we also recommend the testing of various masking processes for actual observation analysis, using both low and high values in order to detect complex MPILs, potentially indicating an imminent eruptive flare activity. 

\section*{Acknowledgements}
This work has received partial support by the European Union's Horizon 2020 Research and Innovation Programme under grant agreement No 640216 (Flare Likelihood and Region Eruption Forecasting [FLARECAST]). JEL is supported by NASA's Heliophysics program. Simulations used in this study were carried out using a grant of computational time from the DoD's High Performance Computing Program. The editor and the authors thank two anonymous referees for their assistance in evaluating this paper and improving its quality.

\clearpage

\end{document}